\documentclass[rmp,aps,preprint,nofootinbib,endfloats]{revtex4}

\usepackage{graphics}
\usepackage{epsfig}

\newcommand{\no}{\noindent}

\newcommand{\beq}{\begin{equation}}
\newcommand{\eeq}{\end{equation}}
\newcommand{\qq}{\begin{eqnarray}}
\newcommand{\qqq}{\end{eqnarray}}

\begin{document}

\title{Exactly solvable Richardson-Gaudin models for many-body quantum
systems}

\date{\today}
\author{J. Dukelsky}
\address{Instituto de Estructura de la Materia, CSIC,
Madrid, Spain}
\author {S. Pittel}
\address{Bartol Research Institute, University of Delaware, Newark, DE
19716 USA}
\author{ G. Sierra}
\address {Instituto de F\'{\i}sica Te\'{o}rica, CSIC/UAM, Madrid, Spain}

\begin{abstract}
The use of exactly-solvable Richardson-Gaudin (R-G) models to
describe the physics of systems with strong pair correlations is
reviewed. We begin with a brief discussion of Richardson's early
work, which demonstrated the exact solvability of the pure pairing
model, and then show how that work has evolved recently into a
much richer class of exactly-solvable models. We then show how the
Richardson solution leads naturally to an exact analogy between
such quantum models and classical electrostatic problems in two
dimensions. This is then used to demonstrate formally how BCS
theory emerges as the large-$N$ limit of the pure pairing
Hamiltonian and is followed by several applications to problems of
relevance to condensed matter physics, nuclear physics and the
physics of confined systems. Some of the interesting effects that
are discussed in the context of these exactly-solvable models
include: (1) the crossover from superconductivity to a
fluctuation-dominated regime in small metallic grains, (2) the
role of the nucleon Pauli principle in suppressing the effects of
high spin bosons in interacting boson models of nuclei, and  (3)
the possibility of fragmentation in confined boson systems.
Interesting insight is also provided into the origin of the
superconducting phase transition both in two-dimensional
electronic systems and in atomic nuclei, based on the
electrostatic image of the corresponding exactly-solvable quantum
pairing models.
\end{abstract}

\maketitle

\newcommand{\bb}{\boldsymbol{\beta}}
\newcommand{\ba}{\boldsymbol{\alpha}}

\tableofcontents

\section{Introduction}
\label{SecI}

Exactly-solvable models have played a major role in helping to
elucidate the physics of strongly correlated quantum systems.
Examples of their extraordinary success can be found throughout
the fields of condensed matter physics and nuclear physics. In
condensed matter, the most important exactly-solvable models have
been developed in the context of one-dimensional (1D) systems,
where they can be classified into three families \cite{Ha}. The
first family began with Bethe's exact solution of the Heisenberg
model \cite{Bet}. Since then a wide variety of 1D models have been
solved using the Bethe ansatz. A second family consists of the
so-called Tomonaga-Luttinger models \cite{Tomo,Lut}, which are
solved by bosonization techniques and which have played an
important role in revealing the non-Fermi-liquid properties of
fermion systems in 1D. For this reason, these systems are now
called Luttinger liquids \cite{Hal}. The third family, proposed by
\citet{Cal} and \citet{Sut} and subsequently generalized to spin
systems \cite{Hal2,Shas}, are models with long-range interactions.
They have been applied to a variety of important problems,
including the physics of spin systems, the quantum Hall effect,
random matrix theory, electrons in 1D, etc.

Exactly-solvable models have also been developed in the field of
nuclear physics, but from a different perspective. In these
models, the Hamiltonian is written as a linear combination of the
Casimir operators of a group decomposition chain chosen to
represent the physics of a particular nuclear phase. One example
is the SU(3) model of \citet{El}, which describes nuclear
deformation and the associated rotational motion. Others include
the three dynamical symmetry limits of the U(6) Interacting Boson
Model \cite{Ia}, which respectively describe rotational nuclei
(the SU(3) limit), vibrational nuclei (the U(5)limit) and
gamma-unstable nuclei (the O(6) limit). These models, all exactly
solvable, have been extremely useful in providing benchmarks for a
description of the complicated collective phenomena that arise in
nuclear systems.

Superconductivity is a phenomenon that is common to both nuclear
physics and condensed matter systems. It is typically described by
assuming a pairing Hamiltonian and treating it at the level of BCS
approximation \cite{BCS}, an approximation that explicitly
violates particle number conservation. While this limitation of
the BCS approximation has a negligible effect for macroscopic
systems, it can lead to significant errors when dealing with small
or ultrasmall systems. Since the fluctuations of the particle
number in BCS are of the order of $\sqrt{N}$ ($N$ being the number
of particles), improvements of the BCS theory are required for
systems with $N\sim 100$ particles. The number-projected BCS
(PBCS) approximation \cite{Mang} has been developed and used in
nuclear physics for a long time and more recently has been applied
in studies of ultrasmall superconducting grains \cite{delft1}. In
the latter, it has been shown necessary to go beyond PBCS
approximation and to resort to the exact solution \cite{duke1} to
properly describe the crossover from the superconducting regime to
the pairing fluctuation regime. It was in this context that the
exact numerical solution of the pairing model, published in a
series of paper in the sixties by \citet{R1,R2,R4,R5,R6} and
\citet{R3} and independently discussed by Gaudin (1995), was
rediscovered and applied successfully to small metallic grains
\cite{random}. [For a review, see \citet{Rep}.] The exact
Richardson solution, though present in the nuclear physics
literature since the sixties, was scarcely used until very
recently, with but a few exceptions \cite{S1, S2, S3}.

Soon after the initial application of the exact solution of the
pairing model to ultrasmall superconducting grains, it became
clear that there is an intimate connection between Richardson's
solution and a different family of exactly-solvable models known
as the Gaudin magnet \cite {Gaudin}. The connection came through
the proof of the integrability of the pairing model by
\citet{CRS}, who identified the complete set of commuting
operators of the model, the quantum invariants whose eigenvalues
are the constants of motion. This made it possible to recast the
pairing Hamiltonian as a linear combination of the quantum
invariants. Furthermore, by establishing a relation between the
constants of motion of the pairing model and those of the Gaudin
magnet, it became possible to generalize the pairing model to
three classes of pairing-like models, all of which were integrable
and all of which could be solved exactly for both fermion and
boson systems \cite{ALO, DES}. The potential importance of these
exactly-solvable models in high $T_c$ superconductivity and other
fields of physics has recently been pointed out \cite{He}. It has
also recently been shown how to obtain the exact solutions of
these models using the Algebraic Bethe ansatz \cite{AFF, ZLMG,
Pog, Lin}, and a connection with Conformal Field Theory and
Chern-Simons theory has been established by \citet{Si} and
\citet{AFS}.

Following their initial discussion of these exactly-solvable
quantum models, \citet{R7} and \citet{G-book} proposed an exact
mapping between these models and a two-dimensional classical
electrostatic problem. By exploiting this analogy, they were able
to derive the thermodynamic limit of the exact solution,
demonstrating that it corresponds precisely to the BCS solution.
Recently, the same electrostatic mapping has been used to provide
a pictorial view of the transition to superconductivity in finite
nuclei  and to suggest an alternative geometrical characterization
of this transition \cite{DEP}. Furthermore, the validity of the
thermodynamic limit based on the electrostatic image has been
numerically checked for very large systems \cite {RSD}.

In this Colloquium, we review the recent progress that has been
made in the use of exactly-solvable pairing and pairing-like
models for the description of strongly-correlated quantum
many-body systems. We begin with a review of the early work of
Richardson and Gaudin, who first showed the exact solvability of
such models. We then discuss the extension to a wider class of
exactly-solvable models, building on the ideas arising from their
integrability. We then discuss several recent applications, to
ultrasmall superconducting grains \cite{random, Sch}, to
interacting boson models of nuclear structure \cite{DP}, to
electrons in a two-dimensional lattice \cite{DES}, and to confined
Bose systems \cite {DS}.

\section{The models of Richardson and Gaudin and their generalization}
\label{SecII}

\subsection{Richardson's exact solution of the pairing model}
\label{SecIIa}

The pairing interaction is the part of the fermion Hamiltonian
responsible for the superconducting phase in metals and in nuclear
matter or neutron stars. It is also responsible for the strong
pairing correlations in the corresponding finite systems, namely
ultrasmall superconducting grains and atomic nuclei. Its
microscopic origin, however, depends on the system under
discussion. In condensed matter, it derives from the exchange of
phonons between the conduction electrons. In nuclear physics, it
comes about due to the short-range nature of the effective
nucleon-nucleon interaction in a nuclear medium and includes
contributions both from the singlet-S and triplet-P channels of
the nucleon-nucleon interaction. The main feature of the pairing
interaction in both cases is that it correlates pairs of particles
in time-reversed states. For recent reviews, see \citet{SU} in
condensed matter and \citet{DH} in nuclear physics.

Quite recently, the first direct sign of BCS superconductivity has
been observed in a trapped degenerate gas of $^{40}K$ fermionic
atoms. The pairing interaction in these trapped dilute systems
comes from the s-wave of the atom-atom interaction \cite{Hei} and
is characterized by the scattering length $a$. This property of
the pairing interaction can be experimentally controlled by
imposing an external magnetic field on the system, thereby tuning
the location of the associated  molecular Feshbach resonance. In
this way, it is possible even to change the sign of the scattering
length, which is a key to producing the observed fermionic
superconductivity.

We begin our discussion of Richardson's solution of the pairing
model by assuming a system of $N$ fermions moving in a set of $L$
single-particle states $l$, each having a total degeneracy $\Omega
_{l}$, and with an additional internal quantum number $m$ that
labels the states within the $l$ subspace.  If the quantum number
$l$ represents angular momentum, the degeneracy of a
single-particle level $l$ is $\Omega _{l}=2l+1$ and $-l\leq m\leq
l$. In general, however, $l$ simply labels different quantum
numbers.  We will assume throughout this discussion of fermion
systems that the $\Omega _{l}$ are even so that for each state
there is another obtained by time-reversal. The operators on which
the pairing Hamiltonian is based are

\begin{equation}
\widehat{n}_{l}=\sum_{m}a_{lm}^{\dagger }a_{lm}\ ,\quad
A_{l}^{\dagger }=\sum_{m}a_{lm}^{\dagger
}a_{l\overline{m}}^{\dagger }=\left( A_{l}\right) ^{\dagger } ~,
\label{ope}
\end{equation}
where $a_{lm}^{\dagger }\left( a_{lm}\right) $ creates
(annihilates) a
particle in the state $\left( lm\right) $ and the state $\left( l\overline{m%
}\right) $ is the corresponding time-reversed state. The number
operator $\widehat{n}_{l}$, the pair creation operator
$A_{l}^{\dagger }$ and the pair annihilation operator $A_{l}$
close the commutation algebra

\begin{equation}
\left[ \widehat{n}_{l},A_{l^{\prime }}^{\dagger }\right] =2\delta
_{ll^{\prime }}~A_{l}^{\dagger },\quad \left[ A_{l},A_{l^{\prime
}}^{\dagger }\right] =2\delta _{ll^{\prime }}\left( \Omega
_{l}-2\widehat{n}_{l}\right) ~.\label{com}
\end{equation}
The corresponding algebra is $SU(2)$.

The most general pairing Hamiltonian can be written in terms of
the three operators in Eq. (\ref{ope}) as

\begin{equation}
H=\sum_{l}\varepsilon _{l}~\widehat{n}_{l}+\sum_{ll^{\prime
}}V_{ll^{\prime }}A_{l}^{\dagger }A_{l^{\prime }} ~. \label{Hgen}
\end{equation}
Often a simplified Hamiltonian is considered, in which the pairing
strengths $V_{ll'}$ are replaced by a single constant $g$, giving
rise to the pairing model (PM) or BCS Hamiltonian,

\begin{equation}
H_{P}=\sum_{l}\varepsilon
_{l}\widehat{n}_{l}+\frac{g}{2}\sum_{ll^{\prime}}
A_{l}^{\dagger}A_{l^{\prime}}  ~. \label{HBCS}
\end{equation}
When $g$ is positive, the interaction is repulsive; when it is
negative, the interaction is attractive.

The approximation leading to the PM Hamiltonian must be
supplemented by a cutoff restricting the number of $l$ states in
the single-particle space. In condensed-matter problems this
cutoff is naturally provided by the Debye frequency of the
phonons. In nuclear physics, the choice of cutoff depends on the
specific nucleus and on the set of active or valence orbits in
which the pairing correlations develop. The cutoff in turn
renormalizes the strength of the effective pairing interaction
that should be used within that active space \cite{Baldo}.

A generic state of $M$ correlated fermion pairs and $\nu $
unpaired particles can be written as

\begin{equation}
\left| n_{1},n_{2},\cdots ,n_{L},\nu \right\rangle =\frac{1}{\sqrt{{\cal N}}}%
\left( A_{1}^{\dagger }\right) ^{n_{1}}\left( A_{2}^{\dagger
}\right) ^{n_{2}}\cdots \left( A_{L}^{\dagger }\right)
^{n_{L}}\left| \nu \right\rangle ~, \label{nanb}
\end{equation}
where ${\cal N}$ is a normalization constant. The number of pairs
$n_l$ in level $l$ is constrained by the Pauli principle to be
$0\leq $ $2n_{l}+\nu _{l}\leq \Omega _{l}$, where $\nu _{l}$
denotes the number of unpaired particles in that level. The
unpaired state $\left| \nu \right\rangle =\left| \nu _{1},\nu
_{2}\cdots \nu _{L}\right\rangle $, with $\nu =\sum_{l}\nu _{l}$,
is defined such that

\begin{equation}
A_{l}\left| \nu \right\rangle =0~,\ \ \widehat{n}_{l}\left| \nu
\right\rangle =\nu _{l}\left| \nu \right\rangle ~. \label{nu}
\end{equation}
A state with $\nu$ unpaired particles is said to have seniority
$\nu$. The total number of collective (or Cooper) pairs is
$M=\sum_{l}n_{l}$ and the total number of particles is $N=2M+\nu
$. \

The dimension of the Hamiltonian matrix in the Hilbert space of
Eq. (\ref{nanb}) quickly exceeds the limits of large-scale
diagonalization for a modest number of levels $L$ and particles
$N$. As an example, consider a problem involving $L$
doubly-degenerate levels and $M=N/2$ pairs. One can readily carry
out full diagonalization for up to $L=16$ and $M=8$, for which the
full dimension of the Hamiltonian matrix  is 12,870. For systems
with up to $L=32$ and $M=16$, the Lanczos method can still be used
to obtain the lowest eigenvalues.  But for larger values of $L$
and $M$, such methods can no longer be used and it is necessary to
find alternative methods to obtain physical solutions. BCS is an
example of an alternative approximate method. Here we focus
instead on finding the exact solution, but without numerical
diagonalization.

In spite of the apparent complexity of the problem, Richardson
showed that the exact unnormalized eigenstates of the hamiltonian
of Eq. (\ref{HBCS}) can be written as

\begin{equation}
\left| \Psi \right\rangle =B_{1}^{\dagger }B_{2}^{\dagger }\cdots
B_{M}^{\dagger }\left| \nu \right\rangle  ~, \label{ansa}
\end{equation}
where the collective pair operators $B_{\alpha }$ have the form
appropriate to the solution of the one-pair problem,

\begin{equation}
B_{\alpha }^{\dagger }=\sum_{l}\frac{1}{2\varepsilon _{l}-E_{\alpha }}%
A_{l}^{\dagger } ~. \label{B}
\end{equation}
In the one-pair problem, the quantities $E_{\alpha }$ that enter
Eq. (\ref{B}) are the eigenvalues of the PM Hamiltonian
(\ref{HBCS}), {\em i.e.}, the {\em pair energies}. Richardson
proposed to use the $M$ pair energies $E_{\alpha}$ in the
many-body wave function of Eq. (\ref{ansa}) as parameters which
are then chosen to fulfill the eigenvalue equation $H_{P}\left|
\Psi \right\rangle =E\left| \Psi \right\rangle $.

We will not repeat here the derivation of the set of equations
that the pair energies must fulfill, referring the reader to
\citet{R3}. In Sect. IV, we will return to the eigenvalue problem
for more general Hamiltonians, of which the pairing Hamiltonian is
a particular case.

The key conclusions from Richardson's derivation are as follows:

\begin{itemize}
\item
The state given in Eq. (\ref{ansa}) is an eigenstate of the PM
Hamiltonian (\ref{HBCS}) if the $M$ pair energies $E_{\alpha }$
satisfy the set of $M$ nonlinear coupled equations (called the
Richardson equations)

\begin{equation}
1-4g\sum_{l}\frac{d_{l}}{2\varepsilon _{l}-E_{\alpha
}}+4g\sum_{\beta \left( \neq \alpha \right) }\frac{1}{E_{\alpha
}-E_{\beta }}=0 ~, \label{Rich}
\end{equation}
where $d_l=\frac{\nu_l}{2}-\frac{\Omega_l}{4}$ is related to the
effective pair degeneracy of single-particle level $l$.

\item
The energy eigenvalue associated with a given solution for the
pair energies is

\begin{equation}
E=\sum_{l}\varepsilon _{l}\nu _{l}+\sum_{\alpha }E_{\alpha } ~.
\label{eig}
\end{equation}
\end{itemize}

Because the Richardson method reduces the problem to solving
non-linear coupled equations,  it can be used for systems well
beyond the limits of either exact numerical diagonalization or the
Lanczos algorithm. For example, the method can be used to obtain
exact solutions for systems with $L=1000$ and $M=500$, for which
the dimension is $2.7\times 10^{299}$. While we cannot obtain all
the solutions for such a problem, we can relatively easily obtain
the lowest few for any value of the coupling constant. This means
that we can study the quantum phase transition for pairing using
the Richardson algorithm but not any phase transitions associated
with temperature. To do the latter, we would need to develop the
Thermodynamic Bethe ansatz for application to the pairing model,
as will be discussed further in the summary given in Sect. VII.

\subsection{The Gaudin magnet}
\label{SecIIb}

Inspired by Richardson's solution of the PM, and building on his
previous work on the Bethe method for solving one-dimensional
problems, \citet{Gaudin} proposed a family of fully integrable and
exactly-solvable spin models. The Gaudin models are based on the
SU(2) algebra of the spin operators,

\begin{equation}
\left[ K^{\alpha },K^{\beta }\right] =2i\varepsilon _{\alpha \beta
\gamma }K^{\gamma } ~, \label{pauli}
\end{equation}
where $K^{\alpha}=\sigma^{\alpha}$ ($\alpha=1,2,3$) are the Pauli
matrices.

A quantum model with $L$ degrees of freedom is integrable if there
exist $L$ independent, global Hermitian operators that commute
with one another. This condition guarantees the existence of a
common basis of eigenstates for the $L$ operators, called the
quantum invariants, and for their eigenvalues, the constants of
motion.

Since the SU(2) algebra has one degree of freedom, the most
general set of $L$ Hermitian operators, quadratic in the spin
operators and global in $L$ spins, are

\begin{equation}
H_{i}=\sum_{j\left( \neq i\right) =1}^{L}\sum_{\alpha
=1}^{3}w_{ij}^{\alpha }~K_{i}^{\alpha }K_{j}^{\alpha } ~,
\label{G1}
\end{equation}
where the $w_{ij}^{\alpha }$ are $3L\left( L-1\right) $  real
coefficients. To define an integrable model, {\it i.e.}, to
satisfy the commutation relations $\left[ H_{i},H_{j}\right] =0$,
these coefficients must satisfy the system of algebraic equations

\begin{equation}
w_{ij}^{\alpha }w_{jk}^{\gamma }+w_{ji}^{\beta }w_{ik}^{\gamma
}-w_{ik}^{\alpha }w_{jk}^{\beta }=0  ~. \label{G2}
\end{equation}

Gaudin proposed two conditions to solve this system of equations.
The first was antisymmetry of the $w$ coefficients,

\begin{equation}
w_{ij}^{\alpha }=-w_{ji}^{\alpha } ~. \label{C1}
\end{equation}
The second was to express the $w$ coefficients as an odd function
of the difference between two real parameters,

\begin{equation}
w_{ij}^{\alpha }=f_{\alpha }\left( \eta _{i}-\eta _{j}\right) ~,
\label{C2}
\end{equation}
in order to fulfill Eq. (\ref{C1}).

The most general solution of Eq. (\ref{G2}) subject to Eqs.
(\ref{C1}) and (\ref{C2}) can be written in terms of elliptic
functions. We further restrict here to the case in which the total
spin in the $z$ direction, $S^{z}=\frac{1}{2}\sum_{i}K_{i}^{z}$,
is conserved, {\it i.e.}, the $L$ operators $H_{i}$ (Gaudin
Hamiltonians) commute with $S^{z}$. Conservation of $S^{z}$
requires that $w_{ij}^{x}=w_{ij}^{y}=X_{ij}$ and
$w_{ij}^{z}=Y_{ij}$, in terms of two new matrices $X$ and $Y$. In
such a case,  the integrability conditions of Eq. (\ref{G2})
reduce to

\begin{equation}
Y_{ij}X_{jk}+Y_{ki}X_{jk}+X_{ki}X_{ij}=0  ~.\label{G3}
\end{equation}

There are three classes of solutions to Eq. (\ref{G3}):
\vspace{0.1in}

\begin{description}
\item [I. The rational model]

\begin{equation}
X_{ij}=Y_{ij}=\frac{1}{\eta _{i}-\eta _{j}}  \label{R}
\end{equation}

\item [II. The trigonometric model]

\begin{equation}
X_{ij}=\frac{1}{\sin \left( \eta _{i}-\eta _{j}\right)
}~,~Y_{ij}=\cot \left( \eta _{i}-\eta _{j}\right)  \label{T}
\end{equation}

\item [III. The hyperbolic model]

\begin{equation}
X_{ij}=\frac{1}{\sinh \left( \eta _{i}-\eta _{j}\right)
}~,~Y_{ij}=\coth \left( \eta _{i}-\eta _{j}\right)  \label{H}
\end{equation}

\end{description}

We will hold off presenting details on the solution of the three
families of Gaudin models until Sect. II.D, where we discuss the
generalization of the Richardson-Gaudin (R-G) models. At that
point we will see that the solutions of the Gaudin models are
based on precisely the same ansatz as used by Richardson in his
solution of the pairing model.

The most general integrable spin hamiltonian that can be written
as a linear combination of the Gaudin integrals of motion in Eq.
(\ref{G1}) is

\begin{eqnarray}
H&=&2\sum_{i}\zeta _{i}H_{i} \nonumber \\
&=&\sum_{i\neq j}\left( \zeta _{i}-\zeta _{j}\right) \left\{
X_{ij}\left[ K_{i}^{x}K_{j}^{x}+K_{i}^{y}K_{j}^{y}\right]
+Y_{ij}K_{i}^{z}K_{j}^{z}\right\} . \nonumber \\
\label{Ham}
\end{eqnarray}

The above Hamiltonian, which  models a spin chain with long-range
interactions, has a total of $2L$ free parameters. There are $L$
$\eta ^{\prime }s$, which define the $X^{\prime}s$, and $L$ $\zeta
^{\prime }s$. It can be readily confirmed that the rational family
gives rise to an $XXX$ spin model while the trigonometric and
hyperbolic families correspond to an $XXZ$ spin model. To the best
of our knowledge, there have been no physical applications of the
Gaudin magnet, though there are indications that when the $2L$
free parameters of the model are chosen at random the Gaudin
magnet behaves as a quantum spin glass \cite{Mar}.

The Richardson model has a natural link to superconductivity and
the Gaudin model to quantum magnetism, both very important
concepts in contemporary physics. Despite these facts, neither
model received much attention from the nuclear or condensed-matter
communities for many years. On the other hand, the Gaudin models
have played an important role in aspects of quantum integrability.
For a recent reference, see \citet{Gould}.

\subsection{Integrability of the pairing model}
\label{SecIIc}

Despite the fact that the Richardson and Gaudin models are so
similar, it was not until the work of \citet{CRS} (CRS) that a
precise connection was established. We now discuss their work and
show how it provides the necessary missing link.

The key point of CRS was to show that the PM is integrable by
finding the set of commuting Hermitian operators in terms of which
the PM Hamiltonian could be expressed as a linear combination.
Finding the complete set of common eigenvectors of those quantum
invariant operators is then equivalent to finding the eigenvectors
of the PM Hamiltonian.

In their derivation, they began by introducing a pseudo-spin
representation of the pair algebra, advancing a connection between
pairing phenomena and spin physics that had been pointed out long
ago by \citet{A}.

The elementary operators of the pair algebra, defined in terms of
the generators of the $SU\left( 2\right) $ pseudo-spin algebra,
are

\begin{eqnarray}
K_{i}^{0}&=&\frac{1}{2}\sum_{m}a_{lm}^{\dagger
}a_{lm}-\frac{1}{4}\Omega _{l}\ ~,  \nonumber \\
K_{l}^{+}&=&\frac{1}{2}\sum_{m}a_{lm}^{\dagger }a_{l\overline{m}%
}^{\dagger }=\left( K_{l}^{-}\right) ^{\dagger }~. \label{K0}
\end{eqnarray}

The operator $K_{l}^{+}$ creates a pair of fermions in
time-reversed states. The degeneracy $\Omega _{l}$  of level $l$
is related to a pseudo-spin $S_l$ for that level according to
$\Omega _{l}=2S_{l}+1$. The three operators in Eq. (\ref{K0})
close the $SU(2$) commutation algebra,
\begin{equation}
\left[ K_{l}^{0},K_{l^{\prime }}^{\pm }\right] =\pm \delta
_{ll^{\prime }}K_{l}^{\pm }~,\quad \left[ K_{l}^{+},K_{l^{\prime
}}^{-}\right] =2\delta _{ll^{\prime }}K_{l}^{0} ~. \label{SU2}
\end{equation}

The $SU\left( 2\right) $ group for a level $l$ has one degree of
freedom and its Casimir operator is

\begin{equation}
\left( K_{l}^{0}\right) ^{2}+\frac{1}{2}\left(
K_{l}^{+}K_{l}^{-}+K_{l}^{-}K_{l}^{+}\right) =\frac{1}{4}\left(
\Omega _{l}^{2}-1\right) ~. \label{casi}
\end{equation}
For a problem involving $L$ single-particle levels, there are
obviously $\ L$ degrees of freedom.

Guided by previous work on the two-level PM, CRS considered the
following set of operators:

\begin{eqnarray}
R_{l}=K_{l}^{0} &+& 2g\sum_{l^{\prime }\left( \neq l\right) }
\frac{1}{\varepsilon _{l}-\varepsilon _{l^{\prime }}}  \left[
\frac{1}{2}\left( K_{l}^{+}K_{l^{\prime
}}^{-}+K_{l}^{-}K_{l^{\prime }}^{+}\right) \right. \nonumber \\
&& ~~~~~~~~+\left. K_{l}^{0} K_{l^{\prime }}^{0} \right] ~.
\label{Rop}
\end{eqnarray}
They showed that these operators are (1) Hermitian, (2) global, in
the sense that they are independent of the Hilbert space, (3)
independent, in the sense that no one can be expressed as a
function of the others, and (4) commute with one another.
Furthermore, there are obviously as many $R_l$ operators as
degrees of freedom. The set of $L$ such operators thus fulfills
the conditions required for the quantum invariants of a fully
integrable model. Finally, they showed that the PM Hamiltonian of
Eq. (\ref{HBCS}) can be written as a linear combination of the
$R_l$ according to

\begin{equation}
H_P=2\sum_{l}\varepsilon _{l}R_{l}+C ~,
\end{equation}
where $C$ is an uninteresting constant.

We now return to the relationship between the Richardson PM and
the Gaudin models. This can be done by focussing on Gaudin's
rational model and comparing its quantum invariants to those of
CRS [see Eq. (\ref{Rop})] for the pairing model. As can be readily
seen, the two are very similar except that the quantum invariants
of Gaudin's rational model are missing a one-body term or
equivalently a linear term in the spin operators. As shown by CRS,
this term preserves the commutability of the $R$ operators and
therefore generalizes Gaudin's rational model. In the following
subsection, we discuss the solution of the so-called generalized
Richardson-Gaudin models that emerge when this term is added.

\subsection{Generalized Richardson-Gaudin models}
\label{SecIId}

The generalization of the Richardson and Gaudin models we now
discuss proceeds along two distinct lines. First, we no longer
limit our discussion to Gaudin's rational model, but now
generalize to all three types (rational, trigonometric and
hyberbolic). We will see that all three can be generalized by the
inclusion of a linear term in the quantum invariants. Second, we
generalize our discussion to include boson models as well as
fermion models.

We begin by considering the generalization to boson models. For
boson systems, the elementary operators of the pair algebra are

\begin{eqnarray}
K_{i}^{0}&=&\frac{1}{2}\sum_{m}a_{lm}^{\dagger
}a_{lm}+\frac{1}{4}\Omega _{l} ~, \nonumber \\
\quad K_{l}^{+}&=&\frac{1}{2}\sum_{m}a_{lm}^{\dagger }a_{l\overline{m}%
}^{\dagger }=\left( K_{l}^{-}\right) ^{\dagger } ~.  \label{KB}
\end{eqnarray}
Note the similarity with the fermion pair operators of Eq.
(\ref{K0}). The only differences are that (1) there is a relative
minus sign between the two terms in the operator $K^{0}$, and (2)
there is no longer a restriction to $\Omega_l$ being even. The
latter point follows from the fact that for bosons a
single-particle state can be its own time-reversal partner. As an
example, consider scalar bosons confined to a 3D harmonic
oscillator potential, as will be discussed further in Sect. VI.D.
The degeneracy associated with a shell having principal quantum
number $n$ is $\Omega_n=(n+1) \cdot (n+2)/2$. The lowest two
shells ($n=0$ and $1$) have odd degeneracies, whereas the next two
($n=2$ and $3$) have even degeneracies

The set of operators in Eq. (\ref{KB}) satisfy the commutation
algebra

\begin{equation}
\left[ K_{l}^{0},K_{l^{\prime }}^{\pm}\right] =\pm\delta
_{ll^{\prime }}K_{l}^{\pm }~,\quad \left[ K_{l}^{+},K_{l^{\prime
}}^{-}\right] =-2\delta _{ll^{\prime }}K_{l}^{0} ~, \label{s11a}
\end{equation}
appropriate to SU(1,1).

In subsequent considerations, we will often treat fermion and
boson systems at the same time. To facilitate this, we  can
combine the relevant SU(2) commutation relations for fermions and
SU(1,1) relations for bosons into the compact form

\begin{equation}
\left[ K_{l}^{0},K_{l^{\prime }}^{\pm}\right] =\pm\delta
_{ll^{\prime }}K_{l}^{\pm }~,\quad \left[ K_{l}^{+},K_{l^{\prime
}}^{-}\right] =\mp 2\delta _{ll^{\prime }}K_{l}^{0} ~. \label{s11}
\end{equation}
When both types of systems are being treated together, we  follow
a convention whereby the upper sign refers to bosons and the lower
sign to fermions.

Following earlier discussion, we now consider the most general
Hermitian and number-conserving operator with linear and quadratic
terms,

\begin{eqnarray}
R_{l}=K_{l}^{0}&+&2g\sum_{l^{\prime }\left( \neq l\right) } \left[
\frac{X_{ll^{\prime }}}{2}  \left( K_{l}^{+}K_{l^{\prime
}}^{-}+K_{l}^{-}K_{l^{\prime }}^{+}\right) \right.
\nonumber\\
&&~~~~~~~~ \mp \left. Y_{ll^{\prime }}K_{l}^{0}K_{l^{\prime }}^{0}
\right]~. \label{Rgen}
\end{eqnarray}
Note that this is a natural generalization of Eq. (\ref{Rop}), but
now appropriate to both boson and fermion systems.

We next look for the conditions that the matrices $X$ and $Y$ in
Eq. (\ref{Rgen}) must satisfy in order that the $R$ operators
commute with one other. Surprisingly, the conditions are precisely
those derived by Gaudin  and presented in Eq. (\ref{G3}), despite
the fact that the quantum invariants now include a linear term and
that they now represent both boson and fermion systems.

Here too three families of solutions derive, which can be written
in compact form as

\begin{equation}
X_{ij}=\frac{\gamma }{\sin \left[ \gamma \left( \eta _{i}-\eta _{j}\right) %
\right] }\ ,\quad Y_{ij}=\gamma \cot \left[ \gamma \left( \eta
_{i}-\eta _{j}\right) \right] ~, \label{XY}
\end{equation}
where $\gamma =0$ corresponds to the rational model, $\gamma =1$
to the trigonometric model and $\gamma =i$ to the hyperbolic
model. The three limits are completely equivalent to those
presented for the Gaudin magnet in Eqs. (\ref{R}-\ref{H}).

The next step is to find the exact eigenstates common to all $L$
quantum invariants given in Eq. (\ref{Rgen}),
\begin{equation}
R_{i}\left| \Psi \right\rangle =r_{i}\left| \Psi \right\rangle ~.
\label{Eq}
\end{equation}
This can be accomplished by using an ansatz similar to the one
used by Richardson [see Eq. (\ref{ansa})] to solve the PM, namely

\begin{equation}
\left| \Psi \right\rangle =\prod_{\alpha =1}^{M}B_{\alpha
}^{\dagger }\left| \nu \right\rangle ~,~B_{\alpha }^{\dagger
}=\sum_{i=1}^{L}u_{i}\left( E_{\alpha }\right) K_{i}^{+} ~.
\label{Anga}
\end{equation}
The wave function in Eq. (\ref{Anga}) is a product of collective
pair operators $B_{\alpha }^{\dagger }$, which are themselves
linear combinations of the raising operators $K_{i}^{+}$ that
create pairs of particles in the various single-particle states.
Note that they are analogous to the Richardson collective pair
operators of Eq. (\ref{B}).

We now present explicitly the solutions to the eigenvalue
equations for the three models. As we will see, it is possible to
present the hyperbolic and trigonometric results in a single set
of compact formulae, by using the notation $sn\,$~for $sin$ or
$sinh$, $cs$ for $cos$ or $cosh$ and $ct$ for $cot$ or $coth$.
These are not to be confused with elliptic functions. The rational
model solutions are of a somewhat different structure and thus are
presented separately. For each family, we first give the
amplitudes $u_i$ that define the collective pairs in terms of the
free parameters $\eta_i$ and the unknown pair energies $E_\alpha$,
then the set of generalized Richardson equations that define the
pair energies $E_\alpha$, and finally the eigenvalues $r_i$ of the
quantum invariants $R_i$.

\begin{description}

\item {\bf I. The rational model}

\begin{equation}
u_{i}\left( E_{\alpha }\right) =\frac{1}{2\eta _{i}-E_{\alpha }}
~, \label{uA}
\end{equation}

\begin{equation}
1\pm 4g\sum_{j}\frac{d_{j}}{2\eta _{j}-E_{\alpha }}\mp
4g\sum_{\beta \left( \neq \alpha \right) }\frac{1}{E_{\alpha
}-E_{\beta }}=0  ~, \label{richA}
\end{equation}

\begin{equation}
r_{i}=d_{i}\left[ 1\mp 2g\sum_{j\left( \neq i\right)
}\frac{d_{j}}{\eta _{i}-\eta _{j}}\mp 4g\sum_{\alpha
}\frac{1}{2\eta _{i}-E_{\alpha }}\right] ~. \label{eigenA}
\end{equation}

\item
{\bf II and III. The trigonometric and hyperbolic models}

\begin{equation}
u_{i}\left( E _{\alpha }\right) =\frac{1}{sn\left( E_{\alpha
}-2\eta _{i}\right) } ~, \label{uB}
\end{equation}

\begin{equation}
1\mp 4g\sum_{j}d_{j}ct(E_{\alpha }-2\eta _{j})\pm 4g\sum_{\beta
\left( \neq \alpha \right) }ct(E_{\beta }-E_{\alpha })=0 ,
\label{richB}
\end{equation}

\[
r_{i}=d_{i}\left\{ 1\mp 2g\left[  \sum_{j\left( \neq i\right)
}d_{j}~ct\left( \eta _{i}-\eta _{j}\right) \right. \right.
\]
\begin{equation}
\left. \left.- 2\sum_{\alpha }ct\left( E_{\alpha }-2\eta
_{i}\right) \right] \right\}  ~. \label{eigenB}
\end{equation}
\end{description}
\ \noindent In Eqs. (\ref{uA}-\ref{eigenB}), the quantity $d_l$ is
now given by
\begin{equation}
d_l ~=~ \frac{\nu_l}{2} ~\pm ~\frac{\Omega_l}{4}~. \label{dlgen}
\end{equation}

Given a set of parameters $\eta_i$ and a pairing strength $g$, the
pair energies $E_{\alpha}$ are obtained by solving a set of $M$
coupled nonlinear equations, either those of Eq. (\ref{richA}) for
the rational model or those of Eq. (\ref{richB}) for the
trigonometric or hyperbolic models. In the limit $g \rightarrow
0$, Eqs. (\ref{richA}, \ref{richB}) can only be satisfied for $
E_{\alpha} \rightarrow 2\eta_i $. In this limit, the corresponding
pair amplitudes $u_{i}(E_\alpha)$ in Eqs. (\ref{uA}, \ref{uB})
become diagonal and the states of Eq. (\ref{Anga}) reduce  to a
product of uncorrelated pairs acting on an unpaired state. The
ground state, for example, involves pairs filling the lowest
possible states and no unpaired particles. Excitations involve
either promoting pairs from the lowest paired states to higher
ones or by breaking pairs and increasing the seniority. In this
way, we can follow the trajectories of the pair energies
$E_{\alpha}$ that emerge from Eqs. (\ref{richA}, \ref{richB}) as a
function of $g$ for each state of the system.

For boson systems the pair energies are always real, whereas for
fermion systems the pair energies can either be real or can arise
in complex conjugate pairs. In the latter case, there can arise
singularities in the solution of Eqs. (\ref{richA}, \ref{richB})
for some critical value of the pairing strength $g_c$, when two or
more pair energies acquire the same value. It was shown by
\citet{R4} that each of these critical $g$ values is related to a
single-particle level $i$ and that at the critical point there are
$1-2 d_i$ pair energies degenerate at $2 \eta_i$. These
singularities of course cancel in the calculations of energies,
which do not show any discontinuity in the vicinity of the
critical points. However, the numerical solution of the nonlinear
equations (\ref{richA}) or (\ref{richB}) may break down for values
of $g$ close to the singularities, making impractical the method
of following the trajectories of the pair energies $E_\alpha$ from
the weak coupling limit to the desire value of the pairing
strength $g$. Recently, \cite{Romb} proposed a new method based on
a change of variables that avoids the singularity problem opening
the possibility of finding numerical solutions in the general
case.

The eigenvalues of the $R$ operators, given by Eqs. (\ref{eigenA})
or (\ref {eigenB}), are always real since the pair energies are
either real or come in complex conjugate pairs. Each solution of
the nonlinear set of equations produces an eigenstate common to
all $R_i$ operators, and consequently to any Hamiltonian that is
written as a linear combination of them. The corresponding
Hamiltonian eigenvalue is the same linear combination of $r_i$
eigenvalues.

\section{The electrostatic mapping of the Richardson-Gaudin
models} \label{SecIII}

We now introduce an exact mapping between the integrable R-G
models and a classical electrostatic problem in two dimensions.
Our derivation builds on the earlier work of \citet{R7} and
\citet{G-book}, who used this electrostatic analogy to show that
the exact solution of the PM Hamiltonian agrees with the BCS
approximation in the large-$N$ limit. For simplicity, we
concentrate here on the exact solution for the rational family
($\gamma=0$). The exact solution of the trigonometric and the
hyperbolic families can be reduced to a set of rational equations
by a proper transformation and then also interpreted as a
classical two-dimensional problem \cite{amic}.

Let us assume that we have a two-dimensional (2D) classical system
composed of a set $M$ free point charges and another set of $L$
fixed point charges. For reasons that will become clear when we
use these electrostatic ideas as a means of studying quantum
pairing problems, we will refer to the fixed point charges as {\it
orbitons} and the free point charges as {\it pairons}.

The Coulomb potential due to the presence of a unit charge at the
origin is given by the solution of the Poisson equation

\begin{equation}
\nabla ^{2}V\left( {\bf r}\right) =-2\pi \delta \left( {\bf
r}\right) ~.\label{Poi}
\end{equation}

The Coulomb potential $V\left( {\bf r}\right) $ that emerges from
this equation depends on the space dimensionality. For one, two
and three dimensions, respectively,  the solutions are

\begin{equation}
V\left( {\bf r}\right) \sim \left\{
\begin{array}{c}
{r} \\
\ln \left( {r}\right)  \\
1/{r}%
\end{array}%
\begin{array}{c}
1D \\
2D \\
3D
\end{array}%
\right.
\end{equation}
In $3D$, the Coulomb potential has the usual $1/r$ behavior, but
in $2D$ it is logarithmic. In practice, there are no $2D$
electrostatic systems. However, there are cases of long parallel
cylindrical conductors for which the end effects can be neglected
so that the problem can be effectively reduced to a $2D$ plane. In
our case, however, we will simply use the mathematical structure
of the idealized 2D problem to obtain useful new insight into the
physics of the quantum many-body pairing problem.

For the purposes of this discussion, we map the 2D $xy$-plane into
the complex plane by assigning to each point ${\bf r}$ a complex
number $z=x+ iy$. Let us now assume that the pairons have charges
$q_{\alpha }$ and positions $z_{\alpha }$, with $\alpha =1,\cdots
,M$, and that the orbitons have charges $q_{i}$ and positions
${z}_{i}$, with $i=1,\cdots ,L $. Since they are confined to a 2D
space, all charges interact with one another through a logarithmic
potential. Let us also assume that there is an external uniform
electric field present, with strength $e$ and pointing along the
real axis. The electrostatic energy of the system is therefore

\begin{eqnarray}
U &=&e\sum_{\alpha =1}^{M}q_{\alpha }{Re(z}_{\alpha })
+e\sum_{j=1}^{L}q_{j}{Re(z}_{j})-\sum_{j=1}^{L}\sum_{\alpha
=1}^{M}q_{\alpha }q_{j}\ln \left\vert {z}_{i}-{z}_{\alpha
}\right\vert   \nonumber  \label{Electro} \\
&&-\frac{1}{2}\sum_{\alpha \neq \beta }q_{\alpha }q_{\beta }\ln
\left\vert {z}_{\alpha }-{z}_{\beta }\right\vert
-\frac{1}{2}\sum_{i\neq j}q_{i}q_{j}\ln \left\vert
{z}_{i}-{z}_{j}\right\vert ~.
\end{eqnarray}

If we now look for the equilibrium position of the free pairons in
the presence of the fixed orbitons, we obtain the extremum
condition

\begin{equation}
e  + \sum_{j} \frac{q_{j}}{z _{j}-z_{\alpha}} - \sum_{\beta \left(
\neq \alpha \right) } \frac{q_{\beta }}{ {z}_{\alpha }-{z}_{\beta
} }=0 ~.\label{Ext}
\end{equation}

It can be readily seen that the Richardson equations for the
rational family (\ref{richA}) coincide with the equations for the
equilibrium position of the pairons (\ref{Ext}), {\it if} the
pairon charges are $q_\alpha = 1$, the pairon positions are
$z_\alpha = E_{\alpha}$, the orbiton charges $q_i$ are equal to
the effective level degeneracies $d_i$ (which for the ground
configuration are $\pm \Omega_i/4$), the orbitons positions are
$z_i = 2 \eta_i$, and the electric field strength is $e = \pm
\frac{1}{4 g}$.

As a reminder, the $\eta_i$'s are free parameters defining the
integrals of motion from which the generalized R-G Hamiltonian is
derived [see Eq. (\ref{C2})]. In the case of the pure PM
Hamiltonian, which is a specific Hamiltonian in the rational
family, they are the single-particle energies of the active
levels. In any subsequent discussion in which they are used, we
will thus refer to them as effective single-particle energies.

Table I summarizes the relationship between the quantum PM and the
classical electrostatic problem implied by the above analogy. From
the above discussion, we see that solving the Richardson equations
for the pair energies $E_{\alpha }$ is completely equivalent to
finding the stationary solutions for the pairon positions in the
analogous classical 2D electrostatic problem.

 \vspace{0.1in}
\begin{center}
TABLE I: Analogy between a quantum pairing problem and the
corresponding 2D classical electrostatic problem.
\end{center}
\begin{center}
\begin{tabular}{|c|c|c|}
\hline Quantum Pairing Model & Classical $2D$ Electrostatic Picture \\
\hline\hline
Effective single particle energy $\eta_i$ & Orbiton position $z_i=2\eta _{i}$ \\
\hline Effective orbital degeneracy $d_i$ & Orbiton charge $q_i=d_i$ \\
\hline Pair energy $E_{\alpha }$ & Pairon position $z_{\alpha}=E_{\alpha }$ \\
\hline Pairing strength $g$ & Electric field strength $e = \pm
\frac{1}{4 g}$ \\
\hline
\end{tabular}
\end{center}

Assuming that the real axis is vertical and the imaginary axis
horizontal, and taking into account that the orbiton positions are
given by the real single-particle energies, it is clear that they
must lie on the vertical axis.  The pairon positions are not of
necessity constrained to the vertical axis, but rather must be
reflection symmetric around it. This reflection symmetry property
can be readily seen by performing complex conjugation on the
electrostatic energy functional (\ref{Electro}). As a consequence,
a pairon must either lie on the vertical axis (real pair energies)
or must be part of a mirror pair (complex pair energies).

The various stationary pairon configurations can be readily traced
back to the weakly interacting system ($g\rightarrow 0$). In this
limit, the pairons for a fermionic system are distributed around
(and very near to) the orbitons, thereby forming compact
artificial {\em atoms}. The number of pairons surrounding orbiton
$i$ cannot exceed $|2 d _{i}|$, as allowed by the Pauli principle,
$i.e.$ we cannot accommodate in a single level more particles than
its degeneracy permits. The lowest-energy (ground-state)
configuration corresponds to distributing the pairons around the
lowest position orbitons consistent with the Pauli constraint. We
then let the system evolve gradually with increasing $g$ until we
reach its physical value. Of course, the Pauli limitation does not
apply to boson systems, which will also be discussed in the
applications.

To illustrate how the analogy applies for a specific quantum
pairing problem involving fermions, we consider the atomic nucleus
$^{114}Sn$. This is a semi-magic nucleus, which can be modelled as
14 valence neutrons occupying the single-particle orbits of the
$N=50-82$ shell. Furthermore, it can be meaningfully treated in
terms of a pure PM Hamiltonian with single-particle energies
extracted from experiment. We will return to this problem briefly
in Sect. VI.B. For now we simply want to use this problem to
illustrate the relationship between the quantum parameters and the
electrostatic parameters for the analogous classical problem.

In Table II, we list the relevant single-particle orbits of the
$50-82$ shell in the third column, their corresponding
single-particle energies in the first column and the associated
degeneracies in the second. Each level corresponds to an orbiton
in the electrostatic problem, with the position of the orbiton
given in the fourth column (at twice the single-particle energy of
the corresponding single-particle level). Note that this is always
pure real, meaning that in the 2D plot each orbiton has $y=0$.
Finally, in the fifth column we give the charge of the orbiton,
which is simply related to the degeneracy of the level according
to the prescription in Table I.

\vspace{0.1in}

\begin{center}
TABLE II: Relationship between the quantum pairing problem for the
nucleus $^{114}Sn$ and the corresponding 2D classical problem. The
notation ``s.p." is shorthand for ``single-particle". The s.p.
energies are in $MeV$.
\end{center}
\begin{center}
\begin{tabular}{|c|c|c|c|c|}
\hline  s.p. energy & s.p.
degeneracy & s.p. level/orbiton & orbiton position& orbiton charge\\
\hline\hline
$0.0$ & $6$ & $d_{5/2}$&  $0.0$ & $-1.5$\\

\hline $0.22$ & $8$ & $g_{7/2}$&  $0.44$ & $-2.0$ \\
\hline $1.90$ &$2$ & $s_{1/2}$ & $3.80$ &$-0.5$\\
\hline  $2.20$ &$4$ &$ d_{3/2}$ & $4.40$ &$-1.0$ \\
\hline $2.80$ & $12$ & $h_{11/2}$ & $5.60$ & $-3.0$\\
\hline\end{tabular}
\end{center}

\begin{figure}
\begin{center}
\includegraphics[width= 8 cm]{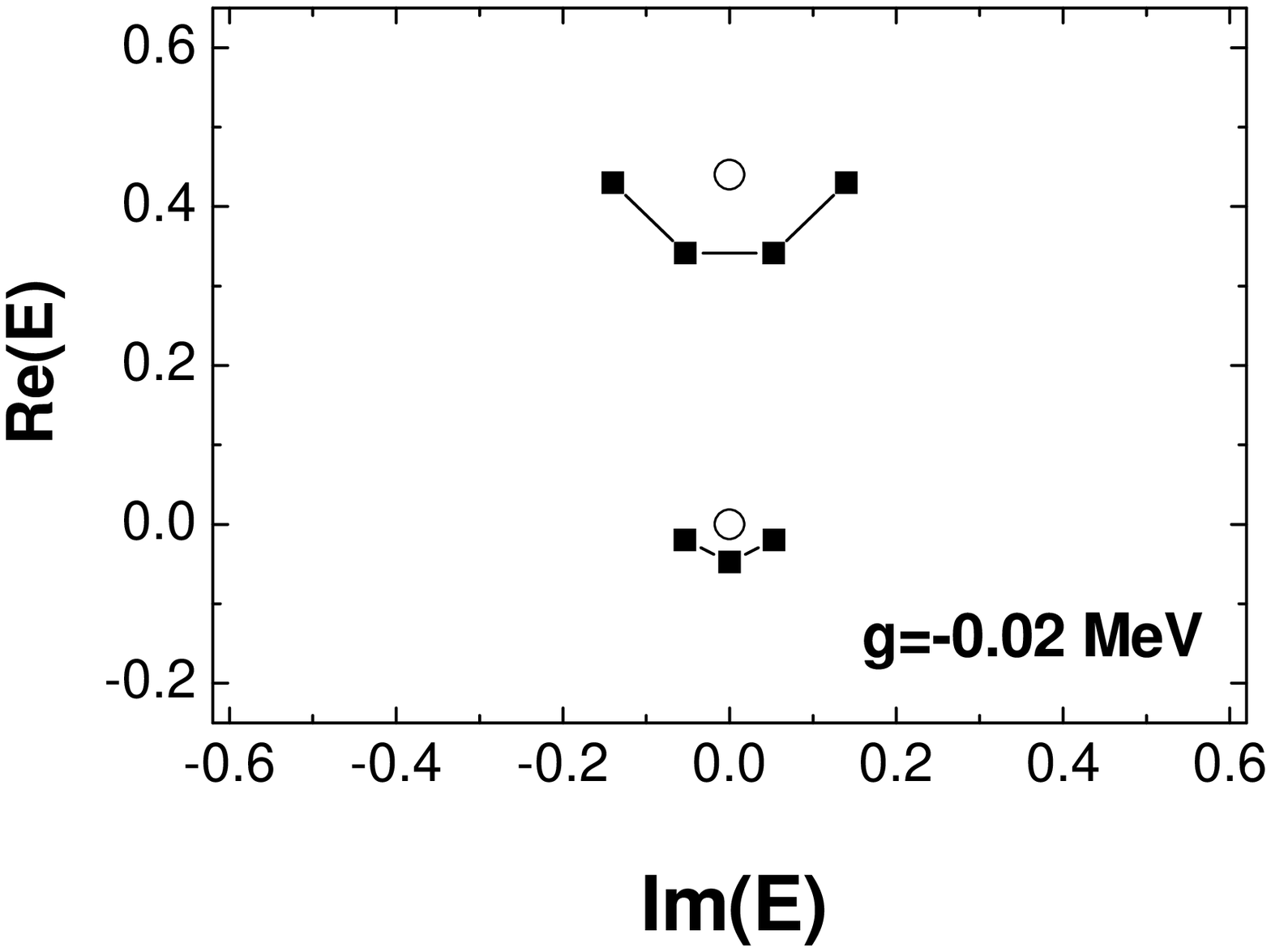}
\end{center}
\caption{Two-dimensional representation of the pairon positions in
$^{114}Sn$ for a pure PM Hamiltonian with single-particle energies
as given in Table II and with $g=-0.02~MeV$. } \label{sn114}
\vspace{1.5cm}
\end{figure}

To illustrate the weak-pairing limit discussed above, we show in
Fig. \ref{sn114} the pairon positions associated with the
electrostatic solution for $^{114}Sn$ calculated for a pairing
strength of $g=-0.02~MeV$, well below the strength at which the
superconducting phase transition sets in. Solid lines connect each
pairon to its nearest neighbor. As we can readily see, the seven
pairons in this case indeed distribute themselves very near to the
lowest two orbitons, with three forming an artificial atom around
the $d_{5/2}$ and four forming an artificial atom around the
$g_{7/2}$.

\section{The large-$N$ limit}
\label{SecIV}

We now discuss how the electrostatic mapping of the previous
section can be used to study the exact solution of the pairing
problem in the large-$N$ or thermodynamic limit. We focus on
fermion systems and consider the PM (or BCS) Hamiltonian used by
\citet{grains} to describe the physics of ultrasmall
superconducting grains,

\beq H_{BCS} = \frac{1}{2}\sum_{j~ \sigma= \pm} \epsilon_{j}~ a_{j
\sigma}^\dagger a_{j \sigma}
  - G \sum_{j ~j'}  a_{j +}^\dagger a_{j -}^\dagger
a_{j' -} a_{j' +} \; , \label{limit1} \eeq

\noindent where $a_{j\pm}$ ($a^\dagger_{j\pm}$) are annihilation
(creation) operators in the time-reversed single-particle states
$|j ~ \pm \rangle$, both with energies
$\varepsilon_j=\epsilon_j/2$, and $G$ is the BCS coupling
constant. Thus, $\epsilon_j$ denotes the energy of a pair
occupying the level $j$ and $\epsilon_{i} \neq \epsilon_{j} $ for
$i \neq j$.

This model was solved analytically by \citet{R3} and numerically
up to $L=32$ single-particle levels by \citet{R5}. The
seniority-zero eigenstates for a system of M fermions depend on a
set of parameters $E_\nu$ ($\nu = 1, \dots , M$) (the pair
energies) that are, in general, complex solutions of the $M$
coupled algebraic Richardson equations

\beq \frac{1}{G } = \sum_{j=1}^L \frac{1}{ \epsilon_j - E_\nu} -
\sum_{\mu=1 (\neq \nu)}^{M} \frac{2}{ E_\mu - E_\nu}\; \qquad \nu
=1, \dots, M ~. \label{limit2} \eeq

\noindent The energy $E$ associated with a given solution is given
by the sum of the resulting pair energies [see Eq. (\ref{eig})].
The ground state is given by the solution of Eq. (\ref{limit2})
with the lowest value of ${E}$.

\begin{figure}
\begin{center}
\includegraphics[width= 6 cm]{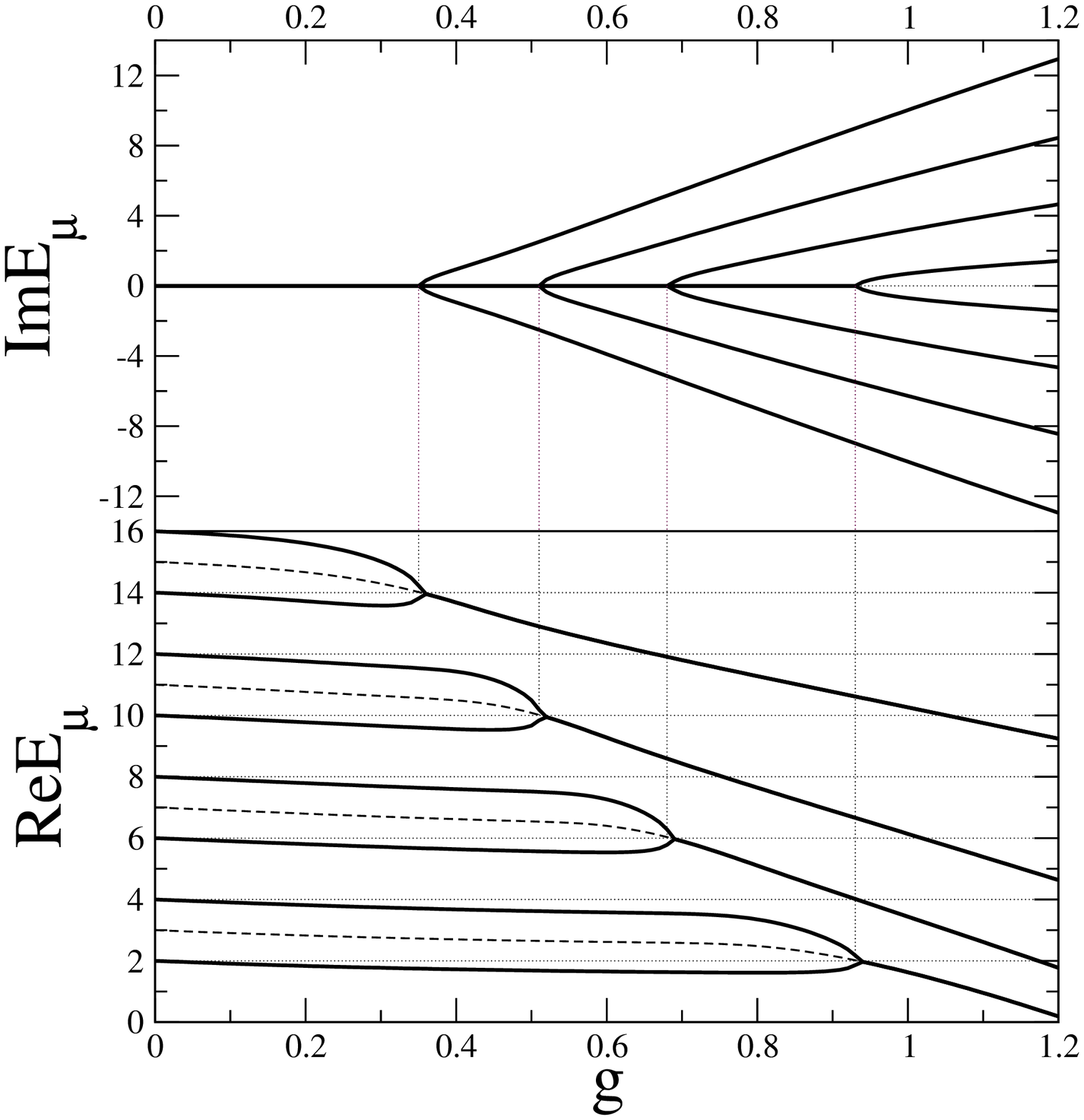}
\end{center}
\caption{Evolution of the real and imaginary parts of $E_\mu(g)$,
in units of the mean level spacing $d = \omega/L$, for the
equally-spaced model with $M=L/2=8$, as a function of the coupling
constant $g$. For convenience, the energy levels are chosen in
this figure as $\epsilon _j = 2 j$. The dashed curves in the lower
half are an average of the two neighboring real energies.
Following the critical point, this turns into the real part of the
energy of the resulting complex conjugate pair.  From
\citet{RSD}.} \label{richenergies8}
\end{figure}

In Fig. \ref{richenergies8}, we plot the solution of Eq.
(\ref{limit2}) for a model of equally-spaced energy levels,
$\epsilon_j = d (2 j - L - 1),\; j=1, \dots,L$, where $d=\omega/L$
is the single-particle level spacing and $\omega$ is twice the
Debye energy. The calculation is done at half filling for
$M=L/2=8$. [Note: At half filling, the number of levels $L$ is
equal to the number of particles $N$.] For small values of the
coupling constant $g = G L$ all the solutions $E_\mu$ are real,
but as we approach some critical value $g_{c,1}$ the two real
roots that are closest to the Fermi level coalesce at $g_{c,1}$
and for $g>g_{c,1}$ develop into a complex conjugate pair. The
same phenomenon happens to other roots as $g$ is further
increased, until eventually all of the roots form complex pairs.
This fact was observed by \citet{R7} and \citet{G-book} and
suggested a way to analyze systems with a large number of
particles, where the exact solution must converge asymptotically
to the BCS solution.

\begin{figure}
\begin{center}
\includegraphics[height= 8 cm,angle= -90]{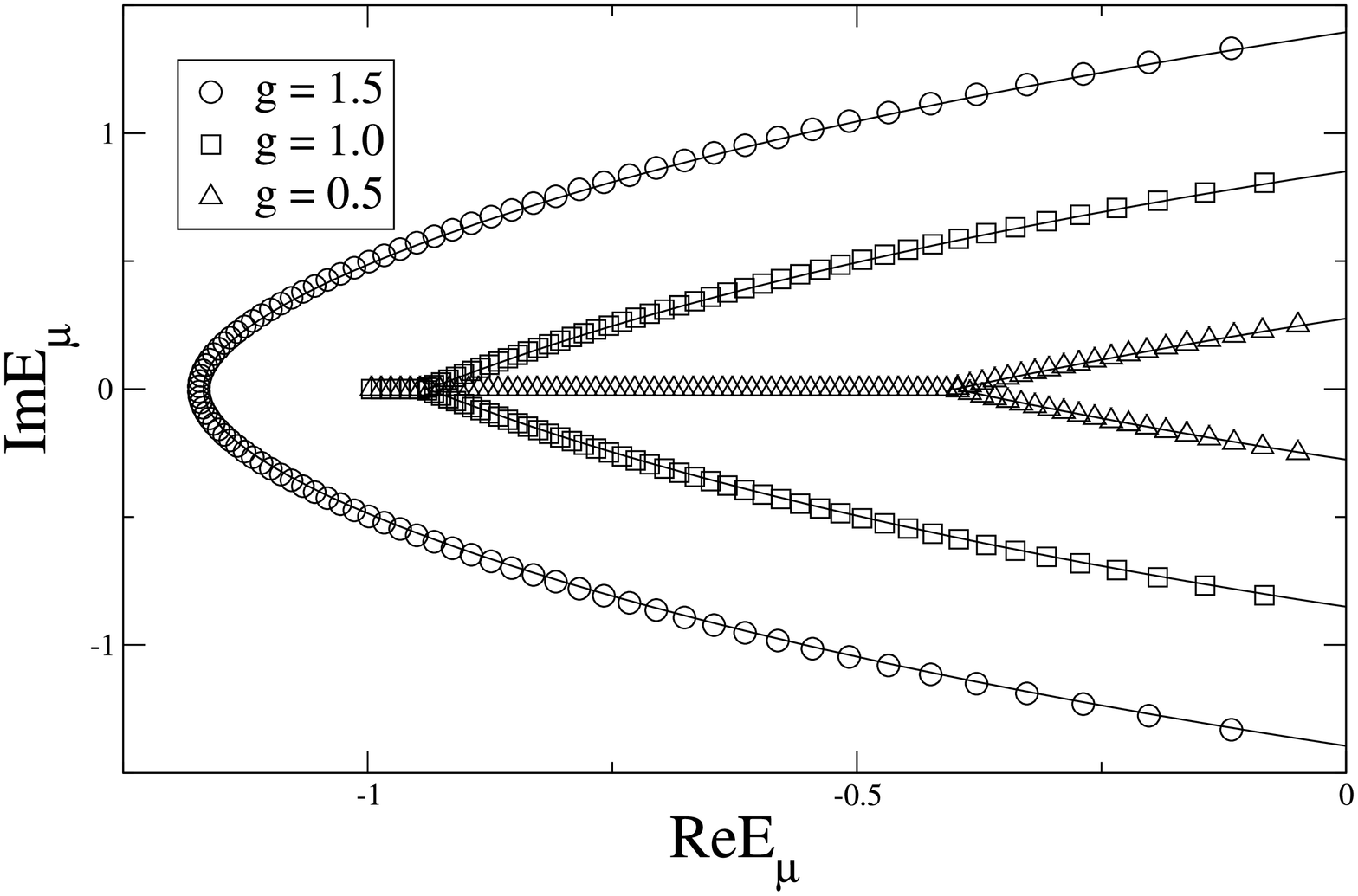}
\end{center}
\caption{Plot of the roots $E_\mu$ for the equally-spaced pairing
model in the complex $\xi$-plane. The discrete symbols denote the
numerical values for $M=100$.  All energies are in units of
$\omega$. From \citet{RSD}.} \label{arcsuniform}
\end{figure}

Figure \ref{arcsuniform} shows the solutions to Eq. (\ref{limit2})
for a system with a much larger number of particles, $M=N/2=100$
pairs, and for three values of $g$. For $g=1.5$, the roots
$E_{\mu}$ form an arc which ends at the points $ 2\lambda \pm
2i\Delta$, where $\lambda$ is the BCS chemical potential and
$\Delta$ the BCS gap. For $g=1.0$ and $0.5$, the set of roots
consist of two pieces, one formed by an arc $\Gamma$ with
endpoints $2\lambda \pm 2i\Delta$ which touches the real axis at
some point $\varepsilon_A$, and a set of real roots along the
segment $[- \omega, \varepsilon_A]$. As $g$ decreases, the latter
segment gets progressively larger while the arc becomes smaller
and eventually shrinks to a point when $g=0$.

The solid lines in the figure are the results obtained from the
algebraic equations derived by \citet{G-book} in the large-$N$
limit. Note that they are in excellent agreement with the results
obtained by numerically solving Eq. (\ref{limit2}).

We now show how to make the connection between the large-$N$ limit
of the PM problem and BCS theory more precise, by applying the
electrostatic analogy introduced in Sect. \ref{SecIII}. We will
consider  the limit in which $L \rightarrow \infty$, while keeping
fixed the following quantities:

\beq G=\frac{g}{L}, \hspace{1cm} \rho=\frac{M}{L} ~. \eeq

Assuming that the pair energies organize themselves into an arc
$\Gamma$ which is piecewise differentiable and symmetric under
reflection on the real axis,  Eq. (\ref{limit2}) (the Richardson
equation) in the continuum limit is converted into the integral
equation

\beq \int_\Omega \frac{ \rho(\epsilon) \; d \epsilon}{\epsilon -
\xi} - P \int_{\Gamma} \frac{ r(\xi') \; |d \xi'|}{ \xi' - \xi} -
\frac{1}{2 G} = 0, \qquad \xi\in \Gamma, \label{limit4} \eeq

\no where $\rho(\epsilon)$ is the energy density associated with
the energy levels that lie in the interval $\Omega = [- \omega,
\omega]$ and satisfies

\beq  \int_\Omega \rho(\epsilon) d \epsilon =  \frac{L}{2}\; ,
\label{limit5} \eeq

\no while $r(\xi)$ is the density of the roots $E_\mu$ that lie in
the arc $\Gamma$ and satisfies

\qq
\int_\Gamma r(\xi) \; |d \xi| & = & M,  \label{limit6} \\
\int_\Gamma \xi \; r(\xi) |d \xi| & = & E. \label{limit7} \qqq

The last equation is a consequence of Eq. (\ref{eig}). The
solution of Eq. (\ref{limit4}) was given by \citet{G-book} using
techniques of complex analysis. We now summarize his main results,
which from a different perspective were also given by \citet{R7}.
A detailed derivation of the continuum limit together with a
comparison with numerical results for large but finite systems was
presented by \citet{RSD}.

 Introducing an ``electric field'', and studying its properties in the vicinity of the arc
$\Gamma$, one can show that Eq. (\ref{limit4}) yields the
well-known BCS gap equation

\beq \int_\Omega  \frac{\rho(\epsilon) \; d\epsilon}{
 \sqrt{(\frac{\epsilon}{2} - \lambda)^2 + \Delta^2}}
 = \frac{1}{G}\; ,
\label{limit11} \eeq

\no that Eq. (\ref{limit6}) becomes the equation for the chemical
potential

\begin{equation}
M = \int_\Omega \left( 1 - \frac{ \frac{\epsilon}{2} - \lambda}{
 \sqrt{(\frac{\epsilon}{2} - \lambda)^2 + \Delta^2}}
\right) \rho(\epsilon)  d\epsilon , \label{limit12}
\end{equation}

\no and that Eq. (\ref{limit7}) gives the BCS expression for the
ground-state energy,

\beq
 E
= - \frac{\Delta^2}{ G} + \int_\Omega \left( 1 - \frac{
\frac{\epsilon}{2} - \lambda}{
 \sqrt{(\frac{\epsilon}{2} - \lambda)^2 + \Delta^2}}
\right) \rho(\epsilon) \; \epsilon \;  d \epsilon.
\label{limit122} \eeq

Thus by using the electrostatic analogy for the quantum pairing
problem we are able to demonstrate how the BCS equations emerge
naturally in the large-$N$ limit.

\section{The elementary excitations of the BCS Hamiltonian}

Most of the studies to date of excited states of the BCS
Hamiltonian [see Eq. (\ref{limit1})] based on Richardson's exact
solution have dealt with a subclass of these states, namely those
obtained by breaking a single Cooper pair. In a case involving
doubly-degenerate levels only, the levels in which the broken pair
reside can no longer be occupied by the collective pairs. Those
levels are thus {\it blocked} and this is reflected in the
Richardson equations by them having effective degeneracy $d_l=0$.
For more general pairing problems, with degeneracies larger than
2, a broken pair will not completely block a single-particle
level, but rather will increase the seniority $\nu_l$ of the level
and give rise to a reduced effective degeneracy
$d_i=\frac{\nu_i}{2}-\frac{\Omega_i}{4}$. There is also another
class of collective excitations that arises without changing the
seniority. These excitations, known in nuclear physics as pairing
vibrations, correspond to the different solutions of the
Richardson equations for a given seniority configuration
${\nu_i}$.

In a mean-field treatment of the same Hamiltonian, the lowest such
excitations of both types correspond to two quasi-particle states
in BCS approximation, which are then mixed by the residual
interaction in Random Phase Approximation (RPA). While this bears
no obvious resemblance to the Richardson approach for these
elementary excitations, it is clear that they should coincide in
the large-$N$ limit.

To investigate this relation, we focus, for simplicity, on the BCS
Hamiltonian of Eq. (\ref{limit1}) and study systematically its
excited states for systems with a fixed number of particles. Once
we know the excited states, we can try to interpret them in terms
of elementary excitations characterized by definite quantum
numbers, statistics and dispersion relations. A generic excited
state will them be given by a collection of elementary
excitations.

\begin{figure}
\begin{center}
\includegraphics[width= 8 cm]{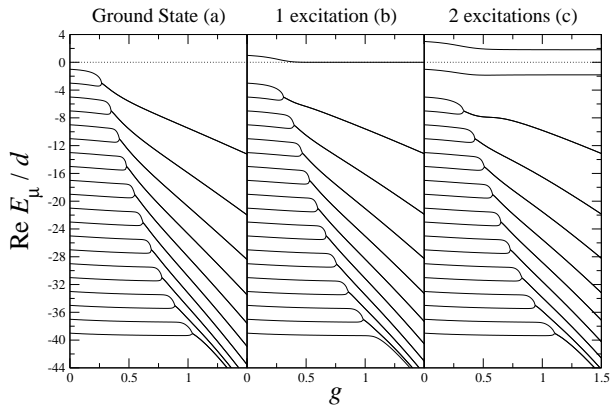}
\end{center}
\caption{Real part of $E_{\mu}$ in units of the mean level spacing
$d$ for the equally-spaced pairing model with $M=L/2=20$
pairs and $N_{G} =0,1,2$ excitations. From \citet{RSD2}.}%
\label{excitfig1}%
\end{figure}

The elementary excitations within the exact Richardson approach
are associated with the number of pair energies $N_G$, either real
or in complex conjugate pairs, that in the large-$g$ limit stay
finite within the interval between neighboring single-particle
levels \cite{RSD2, Yuz}. We display this behavior in Fig.
\ref{excitfig1} for the ground state (GS)  and the first two
excited states for a system with $L=40$ single-particle levels at
half filling ($M=20$) as a function of $g$. Note that for the
ground state all pair energies become complex and their real parts
go to infinity as the coupling strength becomes infinite. Thus
$N_G=0$ for this state. Correspondingly we find that for the first
and second excited states the number of pair energies that remain
finite are $N_G=1$ and $2$, respectively.

In the large-$g$ limit, the pair energies that stay finite ($E^{f}
_\nu$) satisfy the Gaudin equation

\begin{equation}
\sum_{j=1}^L \frac{1}{ \epsilon_j - E^f _\nu} ~-~ \sum_{\mu=1
(\neq \nu)}^{N_G} \frac{2}{ E^f _\mu - E^f _\nu} =0 \;, \qquad \nu
=1, \dots, N_G, \label{limit13} \end{equation} while the remaining
$M-N_G$ pair energies ($E^{i} _\nu$) go to infinity and are the
solution of the generalized Stieltjes  problem encountered by
\citet{Shas2} in the study of the excitations of the ferromagnetic
Heisenberg model,

\begin{equation}
\frac{1}{g}+ \frac{L}{ E^i _\nu} + \sum_{\mu=1 (\neq \nu)}^{M-N_G}
\frac{2}{ E^i _\mu - E^i _\nu} =0 \;, \qquad \nu =1, \dots,
M-N_G~. \label{limit14} \end{equation}

\begin{figure}
\begin{center}
\includegraphics[width= 6 cm]{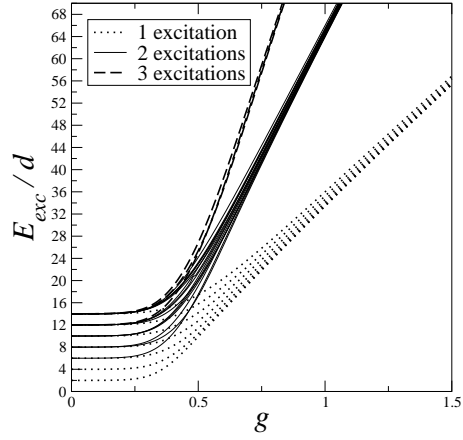}
\end{center}
\caption{Lowest $44$ excitation energies $E_{exc}=E - E_{GS}$ in
units of the mean level spacing $d$ for $M= 20$ pairs
at half filling. From \citet{RSD2}.}%
\label{excitfig5}%
\end{figure}

The fact that the elementary excitations are related to the
trapped pair energies can be readily seen in Fig. \ref{excitfig5},
where we show the low-lying excited states for the same system as
in Fig. \ref{excitfig1}. We see clear evidence in the figure of a
phase transition, which takes place at roughly $g \sim 0.3$. For
lower values of $g$, the states of the system are classified by
the single-particle configurations.  After the transition, this is
no longer the case. We claim that this transition is from a normal
(essentially uncorrelated) system (at small $g$) to a strongly
correlated superconducting system (for large $g$). That crossings
take place around the transition region is a unique characteristic
of an integrable model \cite{Ari}. Observation of level repulsion
in the spectrum would immediately signal non-integrability.

In the extreme superconducting limit ($g\rightarrow\infty$), the
states with the same number $N_G$ of excitations are degenerate.
Moreover, the slope of the excitation energies in that limit is
given by $N_G$.

The degeneracies of the states in the extreme superconducting
limit $d_{L,M,N_{G}}$ have been obtained by Gaudin using the fact
that the Richardson model maps onto the Gaudin magnet in this
limit, and are given by

\begin{equation}
d_{L,M,N_{G}} = C^{L}_{N_{G}} - C^{L}_{N_{G} -1}, \quad0 \leq
N_{G} \leq
M ~,\label{exc1}%
\end{equation}
where $C^L_N$ is the combinatorial number of $N$ permutations of
$L$ numbers. They  satisfy the sum rule $C^{L}_{M} =
\sum_{N_{G}=0}^{M} \; d_{L,M,N_{G}}$, so that the sum of
degeneracies is the total degeneracy.

\begin{figure}
\begin{center}
\includegraphics[width= 8 cm,angle= 0]{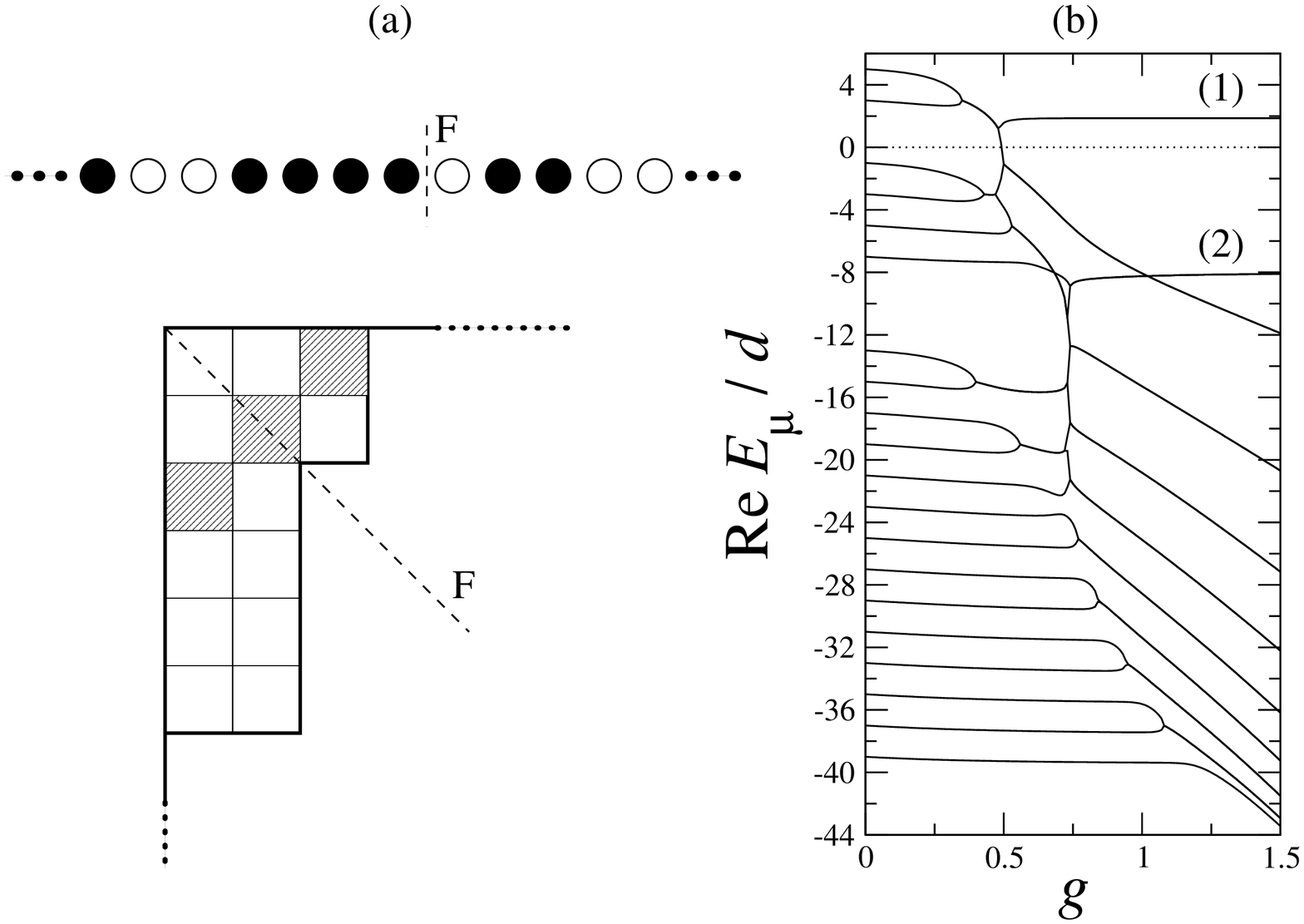}
\end{center}
\caption{An illustration of the algorithm for determining the
number of roots $N_G$, for a case with $N_G=3$. a) Top: Initial
state at $g=0$, where each solid circle ($\bullet$) denotes a
level occupied by a pair, each empty cicle  ($\circ$) denotes an
empty level, and the vertical dashed line with an F near it
denotes the position of the Fermi surface. Bottom: The associated
Young diagram.  b) Real part of $E_\mu$. From \citet{RSD2}.}
\label{excitfig3}
\end{figure}

In general, the practical way to solve the Richardson equations is
to start with a given configuration at $g=0$ and to let the system
evolve with increasing $g$.  Hence the problem is to find for each
initial state the number of roots $N_{G}$ that remain finite in
the $g \rightarrow\infty$ limit. This is a highly non-trivial
problem as it connects the two extreme cases of $g=0$ and
$g=\infty $. The algorithm that relates each unperturbed
configuration to $N_{G}$ has been worked out by \citet{RSD2} in
terms of Young diagrams. We will not give further details here,
but show in Fig. \ref{excitfig3} a particular example with $N_{G}
= 3$ to illustrate the complexity of the evolution of the real
part of the pair energies as a function of $g$. For sufficiently
large values of $g$ and after a complicated pattern of fusion and
splitting of roots (2 real roots $\leftrightarrow$ 1 complex
root), the final result of $N_{G}=3$ emerges.

These results show the non-trivial nature of the elementary
excitations of the pairing model, as exemplified by their
non-trivial counting. As noted above, the elementary excitations
satisfy an effective Gaudin equation. They also satisfy a
dispersion relation similar to that of Bogolioubov quasiparticles.
For these reasons, this new type of elementary excitation has been
called {\em gaudinos} by  \citet{RSD2}. An interesting problem,
which has been partially addressed by \citet{Yuz}, would be to
analyze in detail the relation between gaudinos and  Bogolioubov
quasiparticles for large systems.

\section{Applications}

\subsection{Ultrasmall superconducting grains}

\citet{Anderson:1959} made the conjecture that superconductivity
must disappear for metallic grains when the mean level spacing
$d$, which is inversely proportional to the volume, is of the
order of the superconducting (SC) gap in bulk, $\Delta$. A simple
argument supporting this conjecture is that the ratio $\Delta/d$
measures the number of electronic levels involved in the formation
of Cooper pairs, so that when $\Delta/d \leq 1$ there are no
active levels accessible to build pair correlations. Apart from
some theoretical studies, this conjecture remained largely
unexplored until the recent fabrication of ultrasmall metallic
grains.

\citet{RBT} (RBT), in a series of experiments, studied the
superconducting properties of ultrasmall Aluminium grains at the
nanoscale. These grains have radii $\sim$ 4-5 nm, mean level
spacings $d \sim 0.45$ mev, Debye energies $\omega_D \sim 34$ mev
and charging energies $E_C \sim 46 $ mev. Since the bulk gap of Al
is $\Delta  \sim 0.38$ mev, this satisfies Anderson's condition,
$d \ge \Delta$, for the possible disappearance of
superconductivity. Moreover the large charging energy $E_C$
implies that these grains have a fixed number of electrons, while
the Debye frequency gives an estimate of the number of energy
levels involved in pairing, namely $\Omega = 2 \omega_D/d \sim
150$, which is rather small. Among other things, RBT found an
interesting parity effect, similar to what is known to occur in
atomic nuclei, whereby grains with an even number of electrons
display properties associated with a SC gap, while the odd grains
show gapless behavior.

These experimental findings produced a burst of theoretical
activity focused on the study of the pairing Hamiltonian [Eq.
(\ref{limit1})] with equally spaced levels, {\em i.e.,}
$\varepsilon_j = j d$. Many different approaches were used to
study this model including: i) BCS approximation projected on
parity  \cite{grains}, ii) number-conserving BCS approximation
(PBCS)  \cite{delft1}, iii) Lanzcos diagonalization with up to
$\Omega =23$ energy levels \cite{Lanczos}, iv) Perturbative
Renormalization Group combined with small diagonalization
\cite{BH}, v) the Density Matrix Renormalization Group (DMRG) with
up to $\Omega = 400$ levels \cite{duke1}, etc. [For a review on
this topic, see \citet{Rep}.] Following this flurry of work, it
was suddenly realized that the pairing model under investigation
had in fact been solved {\it exactly} \`a la Bethe by Richardson
long before. This came as a surprise and led to a posteriori
confirmation of the results obtained by the ``exact" numerical
methods, namely Lanzcos Diagonalization and the DMRG. Moreover,
the rediscovery of the Richardson solution produced other
important new developments, including generalization of its
solution, new insight from the point of view of integrable vertex
models, connection with Conformal Field Theory, Chern-Simons
Theory, etc. [For a review, see \citet{Scomo}].

Returning to the application of the Richardson solution to
ultrasmall superconducting grains, we shall focus on two
quantities, the condensation energy and the Matveev-Larkin
parameter \cite{Matveev}. The condensation energy is the
difference between the ground state energy of the pairing
Hamiltonian and the energy of the Fermi state (FS), namely the
Slater determinant obtained by simply filling all levels up to the
Fermi surface. It is given by

\beq E^C_b = E_b^{GS} - \langle FS| H_{BCS} | FS \rangle ~,
\label{grains1} \eeq

\no where $b=0$ for even-parity  grains and $b=1$ for those with
odd parity. In the BCS solution, appropriate when the number of
electrons $N$ is very large, the leading-order behavior of $E^C_b$
is given by $- \tilde {\Delta}^2/(2d)$, where $\tilde{\Delta}$ is
the BCS gap in bulk and $d$ scales as $1/N$, suggesting that
$E^C_b$ is independent of the parity $b$ of the system. However,
an odd ultrasmall grain has a single electron occupying the level
nearest to the Fermi energy. One can easily show that this
electron decouples from the dynamics of the pairing Hamiltonian,
since the pairing interaction only scatters pairs from energy
levels that are doubly occupied to those that are empty. Hence the
single electron only contributes through its free energy.
Furthermore, since there is one less active level at the Fermi
energy, it is harder for the pairing interaction to overcome the
gap and the total energy thus increases. This is the physical
origin of the parity effect in superconducting grains.

Recall that the BCS gap in bulk is given by $\tilde{\Delta} = d
N/\sinh(1/g)$, with $g=0.224$ for Aluminum grains. The BCS result
is obtained by solving the gap equation with a finite number of
energy levels $N$. For even grains, there is a critical value of
the ratio $d^c_0/\tilde{\Delta}= 3.53$, above which there is no
solution to the gap equation. If the grains are odd,  the singly
occupied (``blocked") level must be eliminated from the
Hamiltonian, and the critical ratio becomes
$d^c_1/\tilde{\Delta}=0.89$. The fact that this is smaller than
the even critical ratio indicates that odd grains are less
superconducting than even grains. Thus BCS provides an explanation
of the parity effect observed by RBT. At the same time, it
suggests the existence of an abrupt crossover between the
superconducting regime and the normal state, as conjectured
originally by Anderson.

However, the BCS ansatz does not have a definite number of
particles, which it only fixes on average. Though irrelevant for a
macroscopic sample, this can be important for systems with a small
number of particles where fluctuations in the phase of the
superconducting order parameter may destroy the superconductivity.
For this reason, \citet{delft1} considered the PBCS state, which
includes number projection and thus does not suffer from this
limitation. There are several important differences between the
results obtained with number projection (PBCS) and without (BCS).
Firstly, the condensation energies $E^C_b$ from the PBCS ansatz
are much lower than those from a corresponding BCS treatment.
Secondly, the sharp transition between the SC regime and the
fluctuation-dominated (FD) regime that arises in BCS is smoothed
out by PBCS. Nevertheless, some BCS features survive the inclusion
of number projection, particularly for odd grains. Lastly, in
contrast to BCS, there is always a solution to the PBCS equations.

\begin{figure}
\includegraphics[width=7 cm]{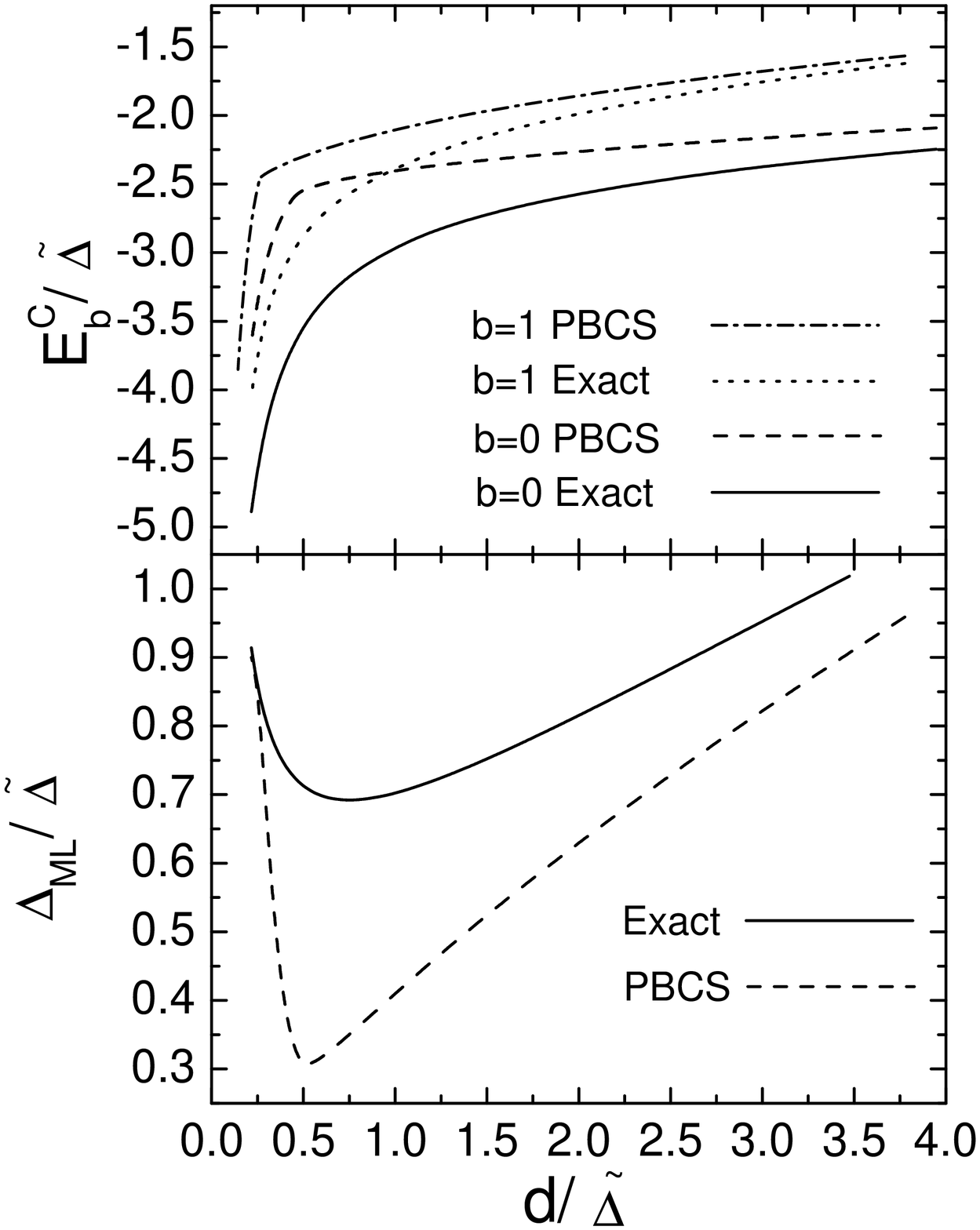}
\caption{ Calculated results for ultrasmall superconducting grains
as a function of the grain size. The upper panel gives the
condensation energies $E^C_b$ for even ($b=0$) and odd $(b=1)$
grains calculated using PBCS and exact wave functions. The lower
panel gives the Matveev-Larkin gap, $\Delta_{ML}$, obtained in
PBCS calculations \cite{delft1} and in exact calculations. The
quantity $\tilde{\Delta}$ that scales the condensation energy, the
Matveev-Larkin gap and the mean level spacing $d$ is the BCS gap
in bulk. } \label{grain} \vspace{2in}
\end{figure}

In the upper panel of Fig. \ref{grain}, we compare the results of
the condensation energy for even and odd grains as a function of
the grain size calculated in the PBCS approximation and exactly.
The exact solution shows a completely smooth SC/FD transition,
although one can still talk about two asymptotic regimes which
match near the level spacing for which the Anderson criterion
$d/\tilde{\Delta} \sim 1$ is satisfied.

Another characterization of the parity effect is in terms of the
gap parameter, which measures the difference between the GS energy
of an odd grain and the mean energy of the neighboring even grains
obtained by adding and removing one electron,

\begin{equation}
\Delta_{ML} = {E}_1(N) - \frac{1}{2} \left( {E}_0(N +1) +
{E}_0(N-1) \right) ~.\label{grains2}
\end{equation}

The lower panel in Fig. \ref{grain} compares the value of $\Delta
_{ML}/\tilde{\Delta} $ computed at the level of PBCS approximation
and exactly. Both curves show a minimum in this quantity as a
function of $d/\tilde{\Delta} $. This latter feature was first
conjectured by Matveev and Larkin (1997) which is why the
associated gap is called the Matveev-Larkin parameter.  It was
subsequently confirmed by \citet{Lanczos} using the Lanczos
method, by \citet{BH} using the Perturbative Renormalization Group
combined with small diagonalization, by \citet{delft1} using the
PBCS method, and by \citet{duke2} using the DMRG method. The shape
of the exact curve, which is identical to that obtained using the
DMRG, is rather smooth as compared with the PBCS method. This can
be interpreted as a suppression of the even-odd parity effect.

Richardson's exact solution of the PM can also be used to study
the interplay of randomness and interactions in a non-trivial
model, by examining the effect of level statistics on the SC/FD
crossover, as reflected for example in the location of the
critical level spacing. There was an earlier study of the latter
question by \citet{SA} using the BCS approach, who concluded that
randomness enhances pairing correlations. More specifically, they
compared the results for a pairing model with a random spacing of
levels (distributed according to a gaussian orthogonal ensemble)
with one having uniform spacings. What they showed is that for
both models the BCS theory gives rise to an abrupt SC/FD
crossover, that the random-spacing Hamiltonian produces a lower
correlation energy $E^C_b$ than the corresponding uniform-spacing
model,  and that the average value of the critical level spacing
in the model based on random splittings is larger than the
corresponding value for the uniform-spacing model. As noted
earlier, however, the mean-field BCS theory produces an abrupt
vanishing of $E^C_b$ that is not present in more sophisticated
treatments. This raises the question of whether the conclusions
they found regarding the role of randomness may be an artifact of
the BCS approach.

\begin{figure}
\begin{center}
\includegraphics[width= 8 cm]{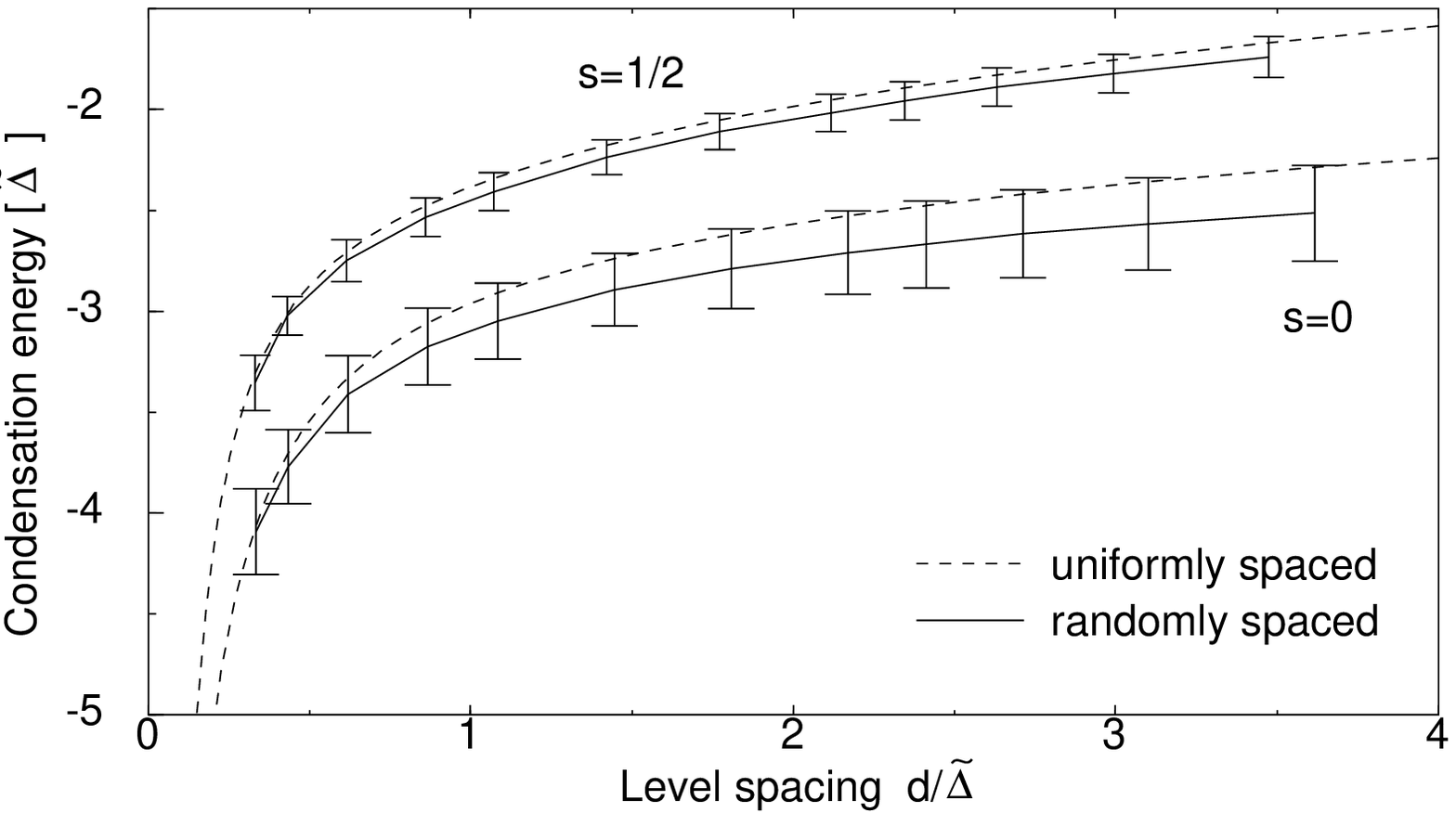}
\end{center}
\caption{ Exact even and odd condensation energies $E^C_b$ for
uniform equally-spaced levels  (dashed line), and the ensemble
average energies $\langle E^C_b \rangle$ for random-spaced levels
(solid line). The height of the fluctuation bars gives the
variances $\delta E^C_b$. The results are plotted as a function of
the mean level spacing, $d$, scaled by the BCS gap in bulk,
$\tilde{\Delta}$. From \citet{random}.}

\label{random}
\end{figure}

Indeed, the exact results shown in Fig. \ref{random} for random
levels show that the SC/FD crossover is as smooth as for the case
of uniformly-spaced levels. This means, remarkably, that even in
the presence of randomness, pairing correlations never vanish, no
matter how large $d/\tilde{\Delta}$ becomes. Quite the contrary,
the randomness-induced lowering of $E^C$ is found to be strongest
in the FD regime.

\subsection{A pictorial representation of pairing in a
two-dimensional lattice}

We next apply the electrostatic analogy to a model of electrons in
a two-dimensional lattice with a residual pairing interaction
\cite{DES}. Assuming a nearest-neighbors hopping term, the
single-electron energies in momentum space are $\varepsilon
_{k}=-2\left( \cos k_{x}+\cos k_{y}\right)$, with $ k_{\sigma
}=2\pi n_{\sigma }/P$ and $-P/2\leq n_{\sigma }<P/2$. In this
expression, $\sigma =x,y$ and $P$ is the number of sites on each
side of the square lattice. In the numerical example that follows,
we consider a $6\times 6$ square lattice at half filling ($ M=18
$) with a constant and attractive pairing Hamiltonian for which
$\varepsilon _{k}=\eta _{k}$. Table III shows the corresponding
information on the positions and charges of the orbitons in the
subspace of seniority-zero states.

\begin{center}
{Table III: Positions and charges of the orbitons for an
attractive
2D pairing model. }\\
\end{center}
\begin{center}
\begin{tabular}{|c|c|c|c|c|c|c|c|c|c|}
\hline $2\varepsilon _{k}$ & $-8$ & $-6$ & $-4$ & $-2$ & $0$ & $2$
& $4$ & $6$ & $8$
\\ \hline\hline
$-\Omega _{k}/4$ & $-1/2$ & $-2$ & $-2$ & $-2$ & $-5$ & $-2$ &
$-2$ & $-2$ & $-1/2$ \\ \hline
\end{tabular}
\end{center}

%%%%%%%%%%%%%%%%%%%%%%%%%%%%%%%%%%%%%%%%%%%%%%%%
\begin{figure}
\begin{center}
\includegraphics[width= 8cm,angle= 0]{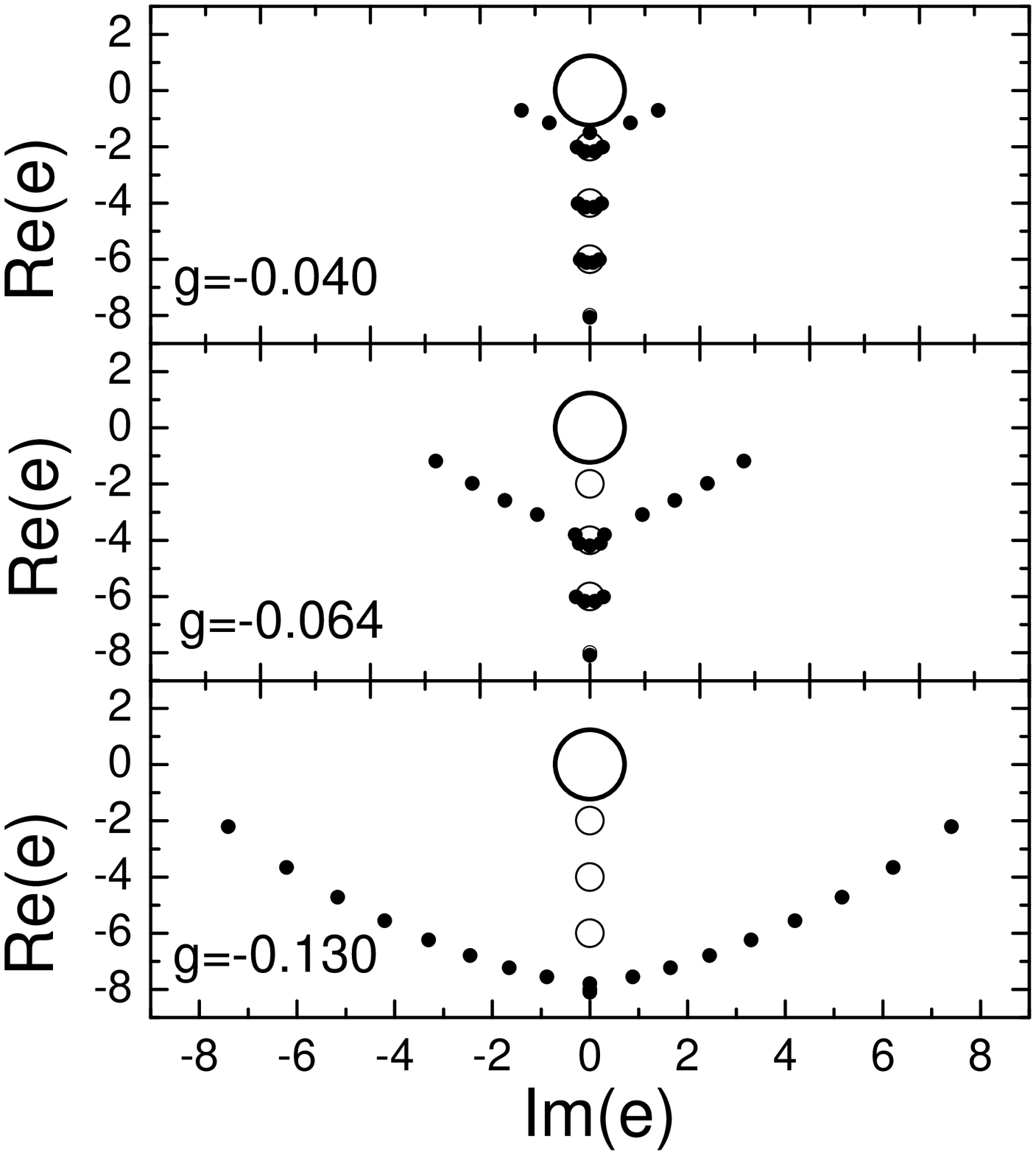}
\end{center}
\caption{Two-dimensional representation of the positions of the
orbitons and pairons corresponding to
 a $6 \times 6$ lattice at half filling, for three selected values of $g$. The
orbitons
 are represented by open circles and the pairons by solid circles, with
 radii
 proportional to their charges.}
\label{2D}
\end{figure}
%%%%%%%%%%%%%%%%%%%%%%%%%%%%%%%%%%%%%%

Figure \ref{2D} shows the equilibrium positions of the pairons
associated with the ground-state solution for three values of $g$.
The orbitons are represented by open circles with radii
proportional to their charges, and the pairons are represented by
solid circles. For each pairon in the figure, we draw a line
connecting it to the one that is closest to it.

In the limit of weak pairing ($g=-0.040$), $5$ pairons are very
near the half-filled orbiton that is located at the Fermi energy
($E=0$) and has charge $-5$. The other $13$ pairons are
distributed close to the lowest four orbitons, consistent with the
Pauli principle. There is one near the lowest and four near each
of the next three. This part of the figure suggests a picture
whereby for weak pairing the pairons organize themselves as
artificial {\em atoms} around their corresponding orbitons. As $g$
increases, the pairons repel, causing the atoms to expand. For
$g=-0.064$ the two orbitons closest to the Fermi energy have lost
their pairons which have now linked up with the five that are near
to the third orbiton. [We can make this linkage more precise by
drawing lines that connect each pairon with its nearest neighbor.]
We refer to this grouping of 13 pairons as a {\em cluster}, as it
arises from the pairons that originally comprised three atoms. The
remaining 5 pairons are still attached to their orbitons, as
artificial atoms. By $g=-0.130$, the cluster has grown to the
point that all 18 pairons are trapped in it.  {\em We claim that
this delocalization effect in the classical problem, from
independent atoms to a collective cluster, is a pictorial
reflection of the transition from a normal to a superconducting
system in the quantum problem.} In the extreme superconducting
limit, as reflected in the figure by the $g=-0.130$ panel, all
pairons are behaving collectively and have lost their memory of
the orbitons from which they arose.  In the corresponding quantum
problem, there are a set of Cooper pairs that likewise behave
collectively and have lost their connection to specific
single-particle orbits.

Similar results have been reported by \citet{DEP} for the problem
of pairing between alike nucleons in atomic nuclei. The analysis
was for two isotopes of $Sn$, including the isotope $^{114}Sn$
briefly alluded to in Sect. III. There too the superconducting
phase transition was seen to be associated with a transition from
isolated atoms to a cluster in the analogous electrostatic
picture. Furthermore, there too the transition to full
superconductivity was seen to develop in steps, depending on the
single-particle levels that play a  role in producing the pair
correlations and their energy hierarchy.

The analysis of pairing in nuclei reported by \citet{DEP} assumed
a pure PM interaction. Of course, this is just an approximation to
the true nuclear interaction in the $J^{\pi}=0^+$ channel.
Nevertheless, we expect that the general features of the
transition to superconductivity should be the same even for a more
realistic pairing interaction. It is in fact possible to build
greater flexibility into the nuclear structure analysis, while
still preserving the electrostatic analogy, by considering more
general exactly-solvable Hamiltonians of the rational family.

\subsection{Electrostatic image of interacting boson models}

As an example of the use of the electrostatic mapping for a finite
boson system, we now discuss the phenomenological Interacting
Boson Model (IBM) of nuclei \cite{Ia}. The IBM captures the
collective dynamics of nuclear systems by representing correlated
pairs of nucleons with angular momentum $L$ by ideal bosons with
the same angular momentum. In its simplest version, known as IBM1,
there is no
distinction between protons and neutrons and only angular momentum $L=0$ ($s$%
) and $L=2$ ($d$) bosons are retained. We will use the
electrostatic image to study the properties of  a second-order
quantum phase transition that arises in the IBM1 from a
vibrational system with $U(5)$ symmetry to a gamma-unstable
deformed system with $O(6)$ symmetry. This phase transition can be
modelled by the one-parameter IBM1 Hamiltonian

\begin{equation}
H=~\widehat{n}_{d}+\frac{x}{2}P^{\dagger }P  ~,\label{hibm}
\end{equation}
where $\widehat{n}_{d}=\sum_{\mu }d_{\mu }^{\dagger }d_{\mu }$,
$P^{\dagger }=s^{\dagger }s^{\dagger }-\sum_{\mu }\left( -1\right)
^{\mu }d_{\mu }^{\dagger }d_{-\mu }^{\dagger }$, $s^{\dagger }$
creates a boson with angular momentum $L=0$, $d^{\dagger}_{\mu}$
creates a boson with angular momentum $L=2$ and z-projection $\mu$
($-2 \le \mu \le 2$), and $x$ is the ratio of the pairing strength
$g$ to the single-particle splitting $\varepsilon_d -
\varepsilon_s$. The parameter $x$ can be varied from $x=0$ (the
$U(5)$ limit) to $x=\infty $ (the $O(6)$ limit). Equation
(\ref{hibm}) is an example of an exactly-solvable repulsive
pairing Hamiltonian, and the second-order nature of the phase
transition it describes has been recently attributed to quantum
integrability \cite{Ari}.

%%%%%%%%%%%%%%%%%%%%%%%%%%%%%%%%%%%%%%%%%%%%%%%%
\begin{figure}
\begin{center}
\includegraphics[width= 7cm,angle= 0]{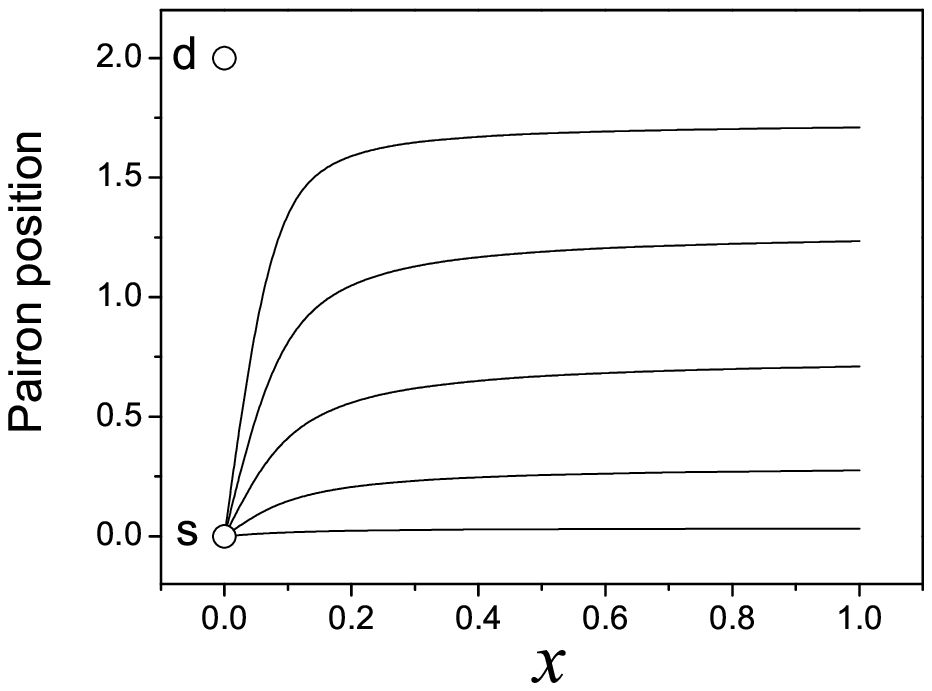}
\end{center}
\caption{Evolution of the pairon positions as a function of the
scaled strength parameter $x$ for a model with $10$ $s$ and $d$
bosons subject to a Hamiltonian with linear single-boson energies
and a repulsive boson pairing interaction. The circles at $x=0$
denote the positions of the $s$ and $d$ orbitons. } \label{fig3}
\end{figure}
%%%%%%%%%%%%%%%%%%%%%%%%%%%%%%%%%%%%%%

The electrostatic problem that corresponds to this quantum boson
model consists of two orbitons with positive charges $q_{s}=1/4$
and $q_{d}=5/4$ (see Sect. III). Both the $s$ and $d$ orbitons are
located on the real axis, with the $s$ located at position $0.0$
and the $d$ at position $2.0$. There are $M$ pairons with positive
unit charge that interact with the orbitons and with one another
and that feel an external electric field pointing downwards with
strength $1/x$. In the ground-state configuration, the pairons are
constrained to move between the two orbitons.

Figure \ref{fig3} shows the pairon positions for a system of $10$
bosons as a function of the control parameter $x$. For $x$ close
to $0$, corresponding to weak repulsive pairing, there is a very
strong electric field, which compresses the pairons very close to
the $s$ orbiton. As $x$ increases, the electric field decreases
and the Coulomb repulsion among all the charges begins to
counteract the effects of the external field. As a consequence,
the pairons gradually expand along the energy interval between the
$s$ and the $d$ orbitons. The phase transition between the
vibrational $U(5)$ system and the gamma-unstable $O(6)$ system,
clearly depicted in the classical electrostatic problem, arises
when the Coulomb repulsion and the external electric field balance
one another. Prior to the phase transition, the quantum system is
primarily an $s$ boson condensate, with perturbative contributions
from $d$ bosons. Following the phase transition, it is a
fragmented condensate mixing the $s$ and $d$ bosons.

A similar analysis by \citet{DP}has also been carried out for a
system with many even angular momentum boson degrees of freedom,
not just the $s$ and $d$. The purpose of that analysis was to
better understand why the IBM1, with just $s$ and $d$ bosons,
works so well in describing the collective properties of nuclei.
Recognizing that bosons in this model are an ideal representation
of the lowest fermion pairs of identical nucleons and that there
are not just $0^+$ and $2^+$ pairs, a natural question to ask is:
Why can we ignore the higher angular momentum pairs/bosons when
dealing with nuclear collective properties? Part of the answer is
contained in the dominant quadrupole-quadrupole neutron-proton
interaction, which is known to favor the lowest $0^+$ and $2^+$
pair degrees of freedom. \citet{DP} suggested another mechanism,
based on an analysis of a generalized boson model containing all
even angular momenta up to some maximum and interacting via a
repulsive boson pairing interaction. The latter is a means of
simulating the repulsive interaction between bosons that arises
due to the Pauli principle between the fermion (nucleon)
constituents of which they are comprised. Using the Richardson
solution of this model, they showed that a repulsive boson pairing
interaction can only correlate two boson degrees of freedom, and
that these should be the lowest two, the $s$ and the $d$. More
recently, this result has been interpreted by means of the
electrostatic mapping \cite{DP2}. Even in the presence of many
boson degrees of freedom and thus many orbitons, the collective
pairons are always confined to lie between the lowest two, {\em
i.e.}, between the $s$ and the $d$.

\subsection{Application to a boson system confined by a harmonic oscillator
trap}

We now consider the problem of a set of bosons confined to a
harmonic oscillator trap and subject to a boson pairing
interaction. We claim that such a Hamiltonian cannot realistically
describe the physics of a confined boson system, for the following
reason.  Looking back at the commutators of the pair operators
$A_{l}^{\dagger }$ given in Eq. (\ref{com}), we see that they are
normalized to the square root of the degeneracy $\Omega _{l}$ of
the level $l$. Thus, the PM Hamiltonian of Eq. (\ref{HBCS}) has a
pairing matrix element proportional to $\sqrt{\Omega _{l}\Omega
_{l^{\prime }}}$. In a three-dimensional harmonic confining
potential, these degeneracies are in turn proportional to $l^{2}$,
where $l$ plays the role of the principal quantum number and the
summation in the pair operators of Eq. (\ref{ope}) now include
both the orbital and the magnetic quantum numbers. On the other
hand, the single-boson energies $\varepsilon _{l}$ for such a
confining potential are linear in $l$. Thus, a boson pairing
interaction in the presence of an oscillator confining trap would
have the net effect of scattering boson pairs to high-lying levels
with greater probability than to low-lying levels, producing
unphysical occupation numbers.

To numerically test this conjecture, we have solved the Richardson
equations [Eq. (\ref{richA})] for a system of 1000 bosons
($M=500$) trapped in a three-dimensional harmonic oscillator
($\Omega _{l}=\left( l+1\right) \left(
l+2\right) /2$ and $\varepsilon _{l}=\hbar \omega (l+3/2)$)with a cutoff at $%
101/2\hbar \omega $ ($L=50$ single boson levels). Following
\citet{R6}, the occupation numbers can be calculated as

\beq \left\langle \widehat{n}_{l}\right\rangle =\left\langle
\frac{\partial H_{P}}{\partial \varepsilon _{l}}\right\rangle
=\sum_{p}\frac{\partial E_{p}}{\partial \varepsilon _{l}}
\label{occup} \eeq

From Eqs. (\ref{HBCS}) and (\ref{richA}), a set of $M$ coupled
nonlinear equations in terms of $M$ new unknowns are obtained,
which when solved give the $L$ occupation numbers. For details of
the derivation , see \citet{R6}.

%%%%%%%%%%%%%%%%%%%%%%%%%%%%%%%%%%%%%%%%%%%%%%%%
\begin{figure}
\begin{center}
\includegraphics[width= 8 cm,angle= 0]{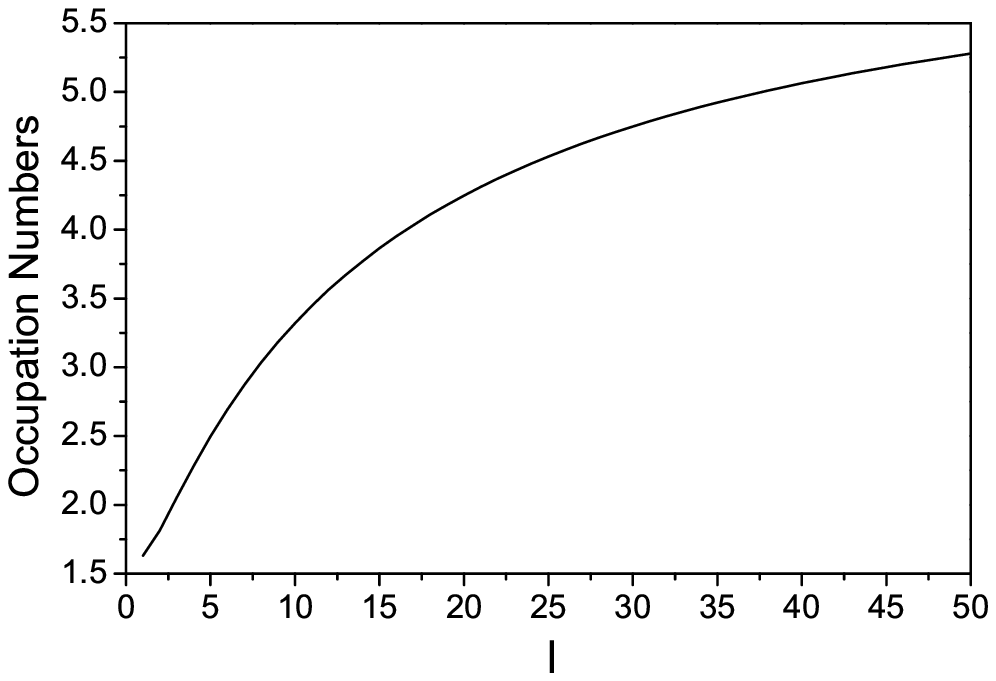}
\end{center}
 \caption{Occupation numbers for $1000$ bosons confined in $50$ harmonic
oscillator shells and interacting via a pure pairing interaction
with strength $g=-.0025$. The occupation of the l=0 state is off
scale and thus not shown.} \label{fig4}
\end{figure}
%%%%%%%%%%%%%%%%%%%%%%%%%%%%%%%%%%%%%%

In Fig. \ref{fig4}, we show the occupation numbers versus the
single-boson energies in units of $\hbar \omega $ for a pairing
strength of $g=-0.0025$. We have excluded the occupation of the
$l=0$ condensed boson state from the figure,  since it lies
outside the scale of the figure. The overall depletion is $0.21$,
which gives an occupation of the $l=0$ state of $n_{0}=790$
bosons. The figure clearly shows that the depletion is
unphysically dominated by the high-lying ({\em i.e.}, high-$l$)
levels, due to the nature of the pair coupling matrix elements
discussed above.

We can use the freedom we have in choosing the parameters $\eta_l
$ entering in the definition of the $R$ operators to obtain a more
physical exactly-solvable model. In order to cancel the unphysical
dependence of the pair coupling matrix elements on the
degeneracies, we choose the $\eta$'s so that $\eta _{l}=\left(
\varepsilon _{l}\right) ^{3}$. Then, the new Hamiltonian, which is
given by the associated linear combination of $R^{\prime }s$, will
be

\begin{equation}
H=2\sum_{l}\varepsilon_{l}R_{l}=C+\sum_{l}\overline{\varepsilon}_{l}n_{l}%
+\sum_{l\neq l^{\prime}}V_{ll^{\prime}}\left[  A_{l}^{\dagger}A_{l^{\prime}%
}-n_{l}n_{l^{\prime}}\right]  ,\label{hrenorm1}%
\end{equation}
where%

\[
C=\frac{1}{2}\sum_{l}\varepsilon_{l}\Omega_{l}-\frac{1}{4}\sum_{l\neq
l^{\prime}}V_{ll^{\prime}}\Omega_{l}\Omega_{l^{\prime}},
\]
\[
~\overline
{\varepsilon}_{l}=\varepsilon_{l}-\sum_{l^{\prime}\left(  \neq
l\right) }V_{ll^{\prime}}\Omega_{l^{\prime}},
\]

\begin{equation}
~V_{ll^{\prime}}=\frac{g}{2}~\frac
{1}{\varepsilon_{l}^{2}+\varepsilon_{l^{\prime}}^{2}+\varepsilon
_{l}\varepsilon_{l^{\prime}}}~.\label{hrenorm2}%
\end{equation}

%%%%%%%%%%%%%%%%%%%%%%%%%%%%%%%%%%%%%%%%%%%%%%%%
\begin{figure}
\begin{center}
\includegraphics [width=8 cm, angle= 0] {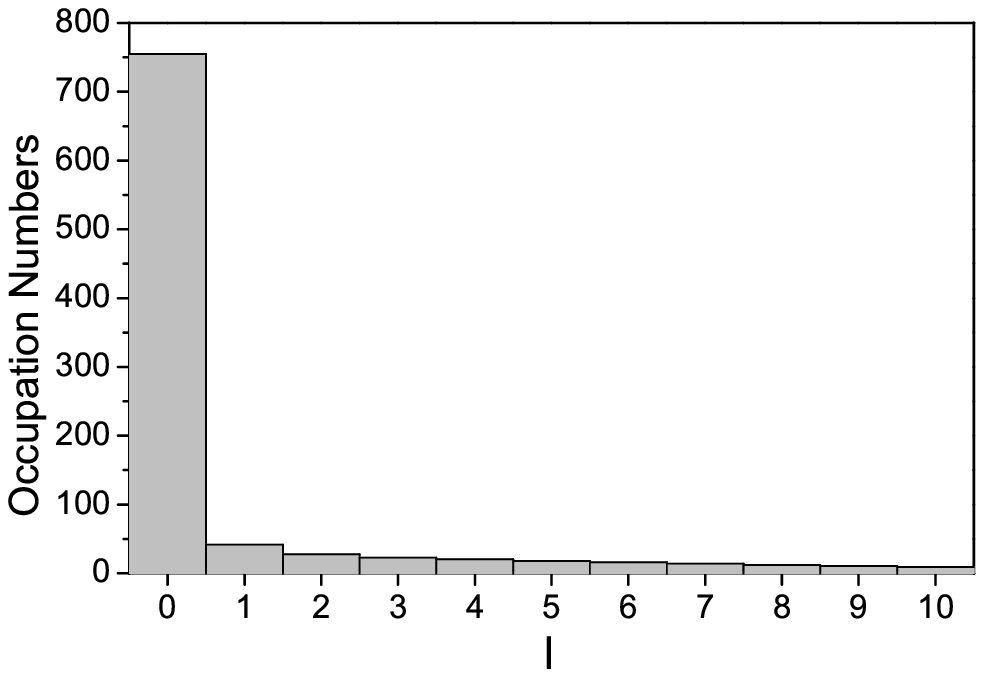}
\end{center}
 \caption{Occupation numbers for $1000$ bosons confined in $50$ harmonic
oscillator shells and interacting via a renormalized pairing
interaction [see Eq. (\ref{hrenorm2})] with $g=-1.0$.}
\label{fig5}
\end{figure}
%%%%%%%%%%%%%%%%%%%%%%%%%%%%%%%%%%%%%%

Taking into account that $\varepsilon _{l}$ is proportional to
$l$, the two-body matrix elements in Eq. (\ref{hrenorm2}) cancel
the dependence on the degeneracies in the effective pair coupling
matrix elements. Thus the above Hamiltonian should be more
appropriate when modelling a harmonically-confined boson system
with a pairing-like interaction.

%%%%%%%%%%%%%%%%%%%%%%%%%%%%%%%%%%%%%%%%%%%%%%%%
\begin{figure}
\begin{center}
\includegraphics[width= 8 cm,angle= 0]{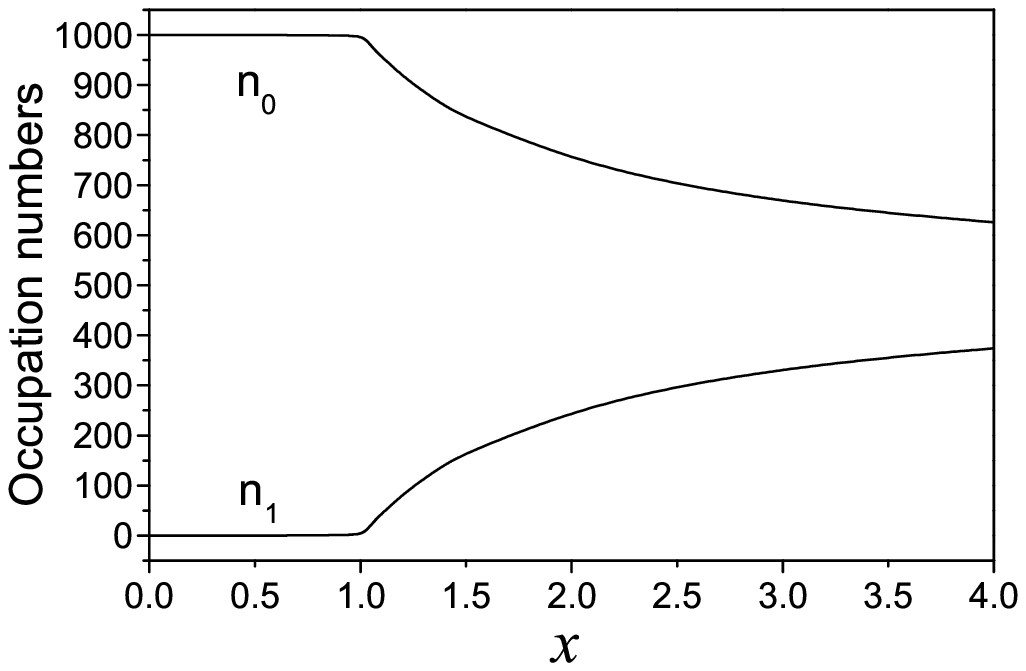}
\end{center}
\caption{Occupation numbers $n_0$ and $n_1$ for $1000$ bosons
confined to $50$ harmonic oscillator shells and interacting via a
repulsive renormalized pairing interaction as a function of the
scaled strength parameter $x$.} \label{fig6}
\end{figure}
%%%%%%%%%%%%%%%%%%%%%%%%%%%%%%%%%%%%%%

The interaction in Eq. (\ref{hrenorm2}) has the nice feature that
its two-body matrix elements decrease with the number of shells,
as one would expect in general. It has the particular property
that the interactions of the pair- and density-fluctuations are
strictly the same but opposite in sign. The energy eigenvalues of
this Hamiltonian can be obtained from the eigenvalues $r_l$ of the
associated $R_l$ operators as $E=2\sum_l \varepsilon_{l}~r_l$,
with the end result being

\begin{equation}
E=\frac{1}{2}\sum_{l}\varepsilon _{l}\Omega
_{l}-\frac{1}{4}\sum_{l\neq
l^{\prime }}V_{ll^{\prime }}\Omega _{l}\Omega _{l^{\prime }}-2g\sum_{lp}%
\frac{\varepsilon _{l}\Omega _{l}}{2\varepsilon _{l}^{3}-E_{p}} ~.
\label{erenorm}
\end{equation}
Note that the first two terms in the energy eigenvalues of Eq.
(\ref{erenorm}) exactly cancel the constant term $C$ in Eq.
(\ref{hrenorm1}).

We have solved the Richardson equations for the Hamiltonian of Eq.
(\ref{hrenorm2}), {\em i.e.,} with $\eta _{l}=\left( \varepsilon
_{l}\right) ^{3}$, for the same system considered above, namely
$M$ =500 and $L$ =50. In Fig. \ref{fig5}, we show the occupation
numbers for $g=-1.0$, for which the overall depletion factor is
$0.245$. They display a reasonable physical pattern, with the
occupancies decreasing monotonically with increasing single-boson
energy. A comparison between the exact results and such
approximations as Hartree-Fock-Bogolyubov theory or its
number-conserving variants will be the subject of future work.

When the same modified form for the Hamiltonian, but with {\em
repulsive} pairing, was considered, a highly unexpected feature
was found \cite{DS}. Figure \ref{fig6} shows the occupation
numbers of the first and second levels versus the scaled pairing
interaction $x={2{M}g}/{\hbar \omega }$ for the same system as
above (${M}=500,\ L=50$). At the critical value of the scaled
interaction, $x_{c}=1,$ the normal ground-state boson condensate
suddenly changes into a new phase in which the bosons occupy the
$l=0$ and the $l=1$ levels, with the occupation of all other
levels negligible.

This new phase is characterized in the large-$N$ limit by having
two macroscopically occupied states, thus representing a
fragmented condensate. It is commonly accepted since the work of
\citet{NSJ} that for confined systems fragmentation cannot occur
in systems of scalar bosons with repulsive interactions. This
might be the first example of fragmentation in a confined boson
system.

\section{Summary and outlook}

In this Colloquium, we have reviewed recent efforts to develop
exactly-solvable models of the Richardson-Gaudin type and have
discussed how these models have been used to provide valuable
insight into the physics of systems with strong pair correlations.
We began with a brief review of Richardson's original treatment of
the pairing model and Gaudin's related treatment of the Gaudin
magnet and then showed how they could be generalized to several
classes of exactly-solvable models that still preserve a
pairing-like structure.

A very attractive feature of these models is that one can
establish an exact analogy between the associated quantum
many-body problem and the classical physics of a two-dimensional
electrostatic system. This feature, which was originally
appreciated by both Richardson and Gaudin, has now been
generalized to all such exactly-solvable models.

The solution to a classical problem is typically amenable to
simple geometrical interpretation, which can then be used to give
an alternative perspective on the quantum model from which it
derives. This has been used in several examples we have discussed,
both for boson and fermion systems. It provides a new perspective
on how superconductivity arises in fermion models and also an
interesting new perspective on a second-order quantum phase
transition that arises in the nuclear Interacting Boson Model.

Another important outcome of the classical electrostatic analogy
is that it facilitates a treatment of these models in the
thermodynamic limit. Not surprisingly, it shows that BCS theory is
indeed the correct large-$N$ limit of the fermion pairing model
with attractive interactions.

The generalization of Richardson and Gaudin's models was first
discussed in the context of the physics of small metallic grains,
as a means of obtaining exact solutions to the BCS Hamiltonian
appropriate to such systems in cases where numerical
diagonalization was out of the question. As discussed in some
detail in this Colloquium, the existence of a method of exact
solution for this model provides important insight into the
detailed physics present when the size of the system gets small
enough so that superconductivity disappears. Neither BCS theory
nor number-projected BCS theory can capture the physics of the
phase transition acceptably.

The new families of exactly-solvable models that we have discussed
are based on the algebras SU(2) (for fermion systems) or SU(1,1)
(for boson systems). These two algebras have a pseudo-spin
representation in terms of fermion pairs or boson pairs,
respectively. All applications to date and all that we have
therefore discussed are based on these representations of the
associated algebras. At the same time, the SU(2) group has other
possible representations in terms of spins operators or two-level
atoms which have not yet been exploited.

Perhaps, the most important feature of the exactly-solvable R-G
models is the enormous freedom within an integrable family. For a
given set of $L$ single-particle levels (or orbits), we can define
an exactly-solvable Hamiltonian in terms of $L+1$ $\eta_i$ and $g$
parameters that enter in the definition of the associated quantum
invariants [Eqs. (\ref{Rgen},\ref{XY})], and an additional set of
$L$ parameters $\varepsilon_i$  that define a Hamiltonian as a
linear combination of the quantum invariants. The models therefore
contain $2L+1$ free parameters, which allows enormous flexibility
in constructing a general pairing-like interaction tailored to the
physical problem of interest \cite{Duke}. In contrast, most other
exactly-solvable models have either no free parameters (the
Heisenberg model), one free parameter (the XXZ model, the Hubbard
model, or the Elliott model), or just a few free parameters (the
three dynamical symmetry limits of the nuclear Interacting Boson
Model). We reported here one particularly interesting use of this
flexibility, namely to model the physics of bosons confined to a
harmonic trap. A pure pairing interaction has the anomalous
feature that it scatters pairs of bosons preferentially to
high-energy states. By exploiting the flexibility of the
generalized R-G models, it was possible to find a more
physically-meaningful Hamiltonian for such systems, which
nevertheless was still exactly solvable. Analysis of this new
model led to a suggestion that confined boson systems can exist in
a fragmented state, contrary to prior expectations.

All the application reported discussed in this Colloquium made use
of generalized R-G models of the so-called rational class. Other
interesting applications within this class of models should
certainly be sought.

A potentially interesting example concerns the properties of the
pairing phase transition in finite Fermi systems at finite
temperature. There has already been important work reported on
this topic (see, for example, \citet{DH} and reference therein).
As noted earlier, the exact solution of the pairing model can be
derived from the Algebraic Bethe ansatz. It is natural, therefore,
to study the Thermodynamic Bethe ansatz, which provides the exact
finite temperature description of short-range $1D$ integrable
models, to see whether it could be extended for application to
integrable pairing models. If so, this would allow for an exact
treatment of the finite temperature properties of such systems as
nuclei, ultrasmall superconducting grains and degenerate Fermi
gases.

Another important topic that has recently received attention
\cite{DH} is the study of the low-energy properties of quantum
many-body systems with random interactions and in particular with
random pairing interactions. Though random pairing interactions in
general have chaotic properties, there is an important subset of
integrable pairing hamiltonians that have a large number of free
parameters which can be chosen randomly. Some of the physical
consequences of randomly chosen pairing interactions in the
general chaotic regime and within the integrable subset of pairing
models have been reported recently by \citet{VZB} and
\citet{Rela}, respectively.

We also expect interesting applications to ensue for the
trigonometric and hyperbolic models. As one example, the
trigonometric model was used by \citet{Gaudin} to derive the limit
of an exactly-solvable model of the interaction of a two-level
atomic system with an external oscillator field. This kind of
model could lead to a generalization of the celebrated
Jaynes-Cummings model. Using techniques similar to the Algebraic
Bethe ansatz derivation of the Richardson model, exactly-solvable
models for boson atomic-molecule systems and two coupled
Bose-Einstein condensates have recently been found \cite{Lin}.
Based on these arguments, we believe that R-G models could have a
promising future in Quantum Optics and in the study of dilute
Fermi and Bose gases.

Another area of great interest is the generalization of R-G models
based on the SU(2) or SU(1,1) algebras to larger algebras. Some
work has already been reported along these lines \cite{AFS}. A
complete solution for models with O(5) or SU(4) symmetries could
lead to interesting applications for nuclear systems with $N
\approx Z$, where it is important to explicitly include the
isospin degree of freedom.  It could also provide useful insight
into the properties of high $T_c$ superconductors \cite{Zhang,Wu}.
Though in previous works \citet{R8,R9} proposed an exact solution
for pairing Hamiltonians with these two group symmetries, recent
work by \citet{Pan} indicates that his solution is invalid for
systems with more than two pairs. Moreover, \citet{o5} and
\citet{su4} have found different solutions for the same problems.
More work is clearly required along these lines.

In closing, the  use of exactly-solvable Richardson-Gaudin models
has already provided significant new and important insight into
the properties of many diverse quantum systems, ranging from
atomic nuclei to electronic systems in condensed matter. We are
optimistic that many more exciting applications are still to come.

\section*{Acknowledgments}

This work was supported in part by the Spanish DGI under grants \#
BFM2000-1320-C02-01/02 and in part by the US National Science
Foundation under Grant Nos. PHY-9970749 and PHY-0140036. We wish
to express our gratitude to Carlos Esebbag, Jos\'{e} Mar\'{\i}a
Rom\'{a}n and Peter Schuck, all of whom contributed significantly
to the work reported in this Colloquium.

\bibliographystyle{apsrmp}

\begin{thebibliography}{299}
\expandafter\ifx\csname natexlab\endcsname\relax\def\natexlab#1{#1}\fi \expandafter\ifx\csname
bibnamefont\endcsname\relax
  \def\bibnamefont#1{#1}\fi
\expandafter\ifx\csname bibfnamefont\endcsname\relax
  \def\bibfnamefont#1{#1}\fi
\expandafter\ifx\csname citenamefont\endcsname\relax
  \def\citenamefont#1{#1}\fi
\expandafter\ifx\csname url\endcsname\relax
  \def\url#1{\texttt{#1}}\fi
\expandafter\ifx\csname urlprefix\endcsname\relax\def\urlprefix{URL }\fi
\providecommand{\bibinfo}[2]{#2} \providecommand{\eprint}[2][]{\url{#2}}



\bibitem[{\citenamefont{Amico \emph{et~al.}}(2001)\citenamefont{Amico, Di Lorenzo, and Osterloh}}]{ALO}
\bibinfo{author}{\bibnamefont{Amico}, \bibfnamefont{L.}},
  \bibinfo{author}{\bibfnamefont{A.}~\bibnamefont{Di Lorenzo}}, and
  \bibinfo{author}{\bibfnamefont{A.}~\bibnamefont{Osterloh}},
  \bibinfo{year}{2001}, \bibinfo{journal}{Phys. Rev. Lett.}
  \textbf{\bibinfo{volume}{86}}, \bibinfo{pages}{5759}.


\bibitem[{\citenamefont{Amico \emph{et~al.}}(2001)\citenamefont{Amico, Falci, and Fazio}}]{AFF}
\bibinfo{author}{\bibnamefont{Amico}, \bibfnamefont{L.}},
  \bibinfo{author}{\bibfnamefont{G.}~\bibnamefont{Falci}}, and
  \bibinfo{author}{\bibfnamefont{R.}~\bibnamefont{Fazio}},
  \bibinfo{year}{2001}, \bibinfo{journal}{J. Phys. A}
  \textbf{\bibinfo{volume}{34}}, \bibinfo{pages}{6425}.



\bibitem[{\citenamefont{Amico \emph{et~al.}}(2002)\citenamefont{Amico, DiLorenzo, Mastellone, Osterloh and Raimondi}}]{amic}
\bibinfo{author}{\bibnamefont{Amico}, \bibfnamefont{L.}},
\bibinfo{author}{\bibfnamefont{A.}~\bibnamefont{Di Lorenzo}},
\bibinfo{author}{\bibfnamefont{A.}~\bibnamefont{Mastellone}},
  \bibinfo{author}{\bibfnamefont{A.}~\bibnamefont{Osterloh}}, and
  \bibinfo{author}{\bibfnamefont{R.}~\bibnamefont{Raimondi}},
  \bibinfo{year}{2002}, \bibinfo{journal}{ Ann. Phys. (N.Y.)}
  \textbf{\bibinfo{volume}{299}}, \bibinfo{pages}{228}.


\bibitem[{\citenamefont{Anderson} (1958)}]{A}
\bibinfo{author}{\bibnamefont{Anderson}, \bibfnamefont{P.~W.}},
  \bibinfo{year}{1958}, \bibinfo{journal}{Phys. Rev.}
  \textbf{\bibinfo{volume}{112}}, \bibinfo{pages}{1900}.


\bibitem[{\citenamefont{Anderson}(1959)}]{Anderson:1959}
\bibinfo{author}{\bibnamefont{Anderson}, \bibfnamefont{P.~W.}},
  \bibinfo{year}{1959}, \bibinfo{journal}{J. Phys. Chem. Solids}
  \textbf{\bibinfo{volume}{11}}, \bibinfo{pages}{28}.

\bibitem[{\citenamefont{Arias \emph{et~al.}}(2003)\citenamefont{Arias, Dukelsky and Garc\'{i}a--Ramos}}]{Ari}
\bibinfo{author}{\bibnamefont{Arias}, \bibfnamefont{J.~E.}},
\bibinfo{author}{\bibfnamefont{J.}~\bibnamefont{Dukelsky}}, and
  \bibinfo{author}{\bibfnamefont{J.~E}~\bibnamefont{Garc\'{\i}a--Ramos}},
  \bibinfo{year}{2003}, \bibinfo{journal}{ Phys. Rev. Lett.}
  \textbf{\bibinfo{volume}{91}}, \bibinfo{pages}{162502}.

\bibitem[{\citenamefont{Arrachea and Rozenberg}(2001)}]{Mar}
\bibinfo{author}{\bibnamefont{Arrachea}, \bibfnamefont{L.}}, and
  \bibinfo{author}{\bibfnamefont{M.~J.}~\bibnamefont{Rozenberg}},
  \bibinfo{year}{2001}, \bibinfo{journal}{Phys. Rev. Lett.}
  \textbf{\bibinfo{volume}{86}}, \bibinfo{pages}{5172}, and private communication.

\bibitem[{\citenamefont{Asorey \emph{et~al}} (2002)\citenamefont{Asorey, DFalceto and Sierra}}]{AFS}
\bibinfo{author}{\bibnamefont{Asorey}, \bibfnamefont{M.}},
  \bibinfo{author}{\bibfnamefont{F.}~\bibnamefont{Falceto}}, and
  \bibinfo{author}{\bibfnamefont{G.}~\bibnamefont{Sierra}},
  \bibinfo{year}{2002}, \bibinfo{journal}{Nucl. Phys. B}
  \textbf{\bibinfo{volume}{622}}, \bibinfo{pages}{593}.


\bibitem[{\citenamefont{Baldo \emph{et~al.}}(1990)\citenamefont{Baldo, Cugnon, Lejeune and Lombardo}}]{Baldo}
\bibinfo{author}{\bibnamefont{Baldo}, \bibfnamefont{M.}},
  \bibinfo{author}{\bibfnamefont{J.}~\bibnamefont{Cugnon}},
  \bibinfo{author}{\bibfnamefont{A.}~\bibnamefont{Lejeune}}, and
  \bibinfo{author}{\bibfnamefont{U.}~\bibnamefont{Lombardo}},
  \bibinfo{year}{1990}, \bibinfo{journal}{Nucl. Phys. A}
  \textbf{\bibinfo{volume}{515}}, \bibinfo{pages}{409}.


\bibitem[{\citenamefont{Bang and Krumlimde}(1970)}]{S1}
\bibinfo{author}{\bibnamefont{Bang}, \bibfnamefont{J.}},
 and  \bibinfo{author}{\bibfnamefont{J.}~\bibnamefont{Krumlimde}},
  \bibinfo{year}{1970}, \bibinfo{journal}{Nucl. Phys. A}
  \textbf{\bibinfo{volume}{141}}, \bibinfo{pages}{18}.


\bibitem[{\citenamefont{Bardeen \emph{et~al.}}(1957)\citenamefont{Bardeen, Cooper and Schrieffer}}]{BCS}
\bibinfo{author}{\bibnamefont{Bardeen}, \bibfnamefont{J.}},
  \bibinfo{author}{\bibfnamefont{L.~N.}~\bibnamefont{Cooper}}, and
  \bibinfo{author}{\bibfnamefont{J.~R.}~\bibnamefont{Schrieffer}},
  \bibinfo{year}{1957}, \bibinfo{journal}{Phys. Rev.}
  \textbf{\bibinfo{volume}{108}}, \bibinfo{pages}{1175}.


\bibitem[{\citenamefont{Berger and Halperin}(1998)}]{BH}
\bibinfo{author}{\bibnamefont{Berger}, \bibfnamefont{S.~D.}},
 and   \bibinfo{author}{\bibfnamefont{B.~I}~\bibnamefont{Halperin}},
  \bibinfo{year}{1998}, \bibinfo{journal}{Phys. Rev. B}
  \textbf{\bibinfo{volume}{58}}, \bibinfo{pages}{5213}.


\bibitem[{\citenamefont{Bethe}(1931)}]{Bet}
\bibinfo{author}{\bibnamefont{Bethe}, \bibfnamefont{H.~A.}},
  \bibinfo{year}{1931}, \bibinfo{journal}{Z. Phys.}
  \textbf{\bibinfo{volume}{71}}, \bibinfo{pages}{265}.

\bibitem[{\citenamefont{Braun and von Delft}(1998)}]{delft1}
\bibinfo{author}{\bibnamefont{Braun}, \bibfnamefont{F.}},
 and   \bibinfo{author}{\bibfnamefont{J.}~\bibnamefont{von Delft}},
  \bibinfo{year}{1998}, \bibinfo{journal}{Phys. Rev. Lett.}
  \textbf{\bibinfo{volume}{81}}, \bibinfo{pages}{4712}.


\bibitem[{\citenamefont{Calogero}(1962)}]{Cal}
\bibinfo{author}{\bibnamefont{Calogero}, \bibfnamefont{F.}},
  \bibinfo{year}{1962}, \bibinfo{journal}{J. Math. Phys.}
  \textbf{\bibinfo{volume}{10}}, \bibinfo{pages}{2191}.

\bibitem[{\citenamefont{Cambiaggio, Rivas and Saraceno}(1997)\citenamefont{Cambiaggio \emph{et~al.}}}]{CRS}
\bibinfo{author}{\bibnamefont{Cambiaggio}, \bibfnamefont{M.~C.}},
  \bibinfo{author}{\bibfnamefont{A.~M.~F.}~\bibnamefont{Rivas}}, and
  \bibinfo{author}{\bibfnamefont{M.}~\bibnamefont{Saraceno}},
  \bibinfo{year}{1997}, \bibinfo{journal}{Nucl. Phys. A}
  \textbf{\bibinfo{volume}{624}}, \bibinfo{pages}{157}.


\bibitem[{\citenamefont{Dean and Hjorth-Jensen}(2003)}]{DH}
\bibinfo{author}{\bibnamefont{Dean}, \bibfnamefont{D.~J.}},
 and  \bibinfo{author}{\bibfnamefont{M.}~\bibnamefont{Hjorth-Jensen}},
  \bibinfo{year}{2003}, \bibinfo{journal}{Rev. Mod. Phys.}
  \textbf{\bibinfo{volume}{75}}, \bibinfo{pages}{607}.


\bibitem[{\citenamefont{Dietrich \emph{et~al.}}(1964)\citenamefont{Dietrich, Mang
  and Pradal}}]{Mang}
\bibinfo{author}{\bibnamefont{Dietrich}, \bibfnamefont{K.}},
  \bibinfo{author}{\bibfnamefont{H.~J}~\bibnamefont{Mang}}, and
  \bibinfo{author}{\bibfnamefont{J.~H}~\bibnamefont{Pradal}},
  \bibinfo{year}{1964}, \bibinfo{journal}{Phys. Rev. B}
  \textbf{\bibinfo{volume}{135}}, \bibinfo{pages}{22}.


\bibitem[{\citenamefont{Dukelsky and Sierra}(1999)}]{duke1}
\bibinfo{author}{\bibnamefont{Dukelsky}, \bibfnamefont{J.}},
 and  \bibinfo{author}{\bibfnamefont{G.}~\bibnamefont{Sierra}},
  \bibinfo{year}{1999}, \bibinfo{journal}{Phys. Rev. Lett.}
  \textbf{\bibinfo{volume}{83}}, \bibinfo{pages}{172}.


\bibitem[{\citenamefont{Dukelsky and Sierra}(2000)}]{duke2}
\bibinfo{author}{\bibnamefont{Dukelsky}, \bibfnamefont{J.}},
 and  \bibinfo{author}{\bibfnamefont{G.}~\bibnamefont{Sierra}},
  \bibinfo{year}{2000}, \bibinfo{journal}{Phys. Rev. B}
  \textbf{\bibinfo{volume}{61}}, \bibinfo{pages}{12302}.


\bibitem[{\citenamefont{Dukelsky \emph{et~al.}}(2001)\citenamefont{Dukelsky, Esebbag
  and Schuck}}]{DES}
\bibinfo{author}{\bibnamefont{Dukelsky}, \bibfnamefont{J.}},
  \bibinfo{author}{\bibfnamefont{C.}~\bibnamefont{Esebbag}}, and
  \bibinfo{author}{\bibfnamefont{P.}~\bibnamefont{Schuck}},
  \bibinfo{year}{2001}, \bibinfo{journal}{Phys. Rev. Lett.}
  \textbf{\bibinfo{volume}{87}}, \bibinfo{pages}{66403}.


\bibitem[{\citenamefont{Dukelsky and Schuck}(2001)}]{DS}
\bibinfo{author}{\bibnamefont{Dukelsky}, \bibfnamefont{J.}}, and
  \bibinfo{author}{\bibfnamefont{P.}~\bibnamefont{Schuck}},
  \bibinfo{year}{2001}, \bibinfo{journal}{Phys. Rev. Lett.}
  \textbf{\bibinfo{volume}{86}}, \bibinfo{pages}{4207}.

\bibitem[{\citenamefont{Dukelsky and Pittel}(2001)}]{DP}
\bibinfo{author}{\bibnamefont{Dukelsky}, \bibfnamefont{J.}}, and
  \bibinfo{author}{\bibfnamefont{S.}~\bibnamefont{Pittel}},
  \bibinfo{year}{2001}, \bibinfo{journal}{Phys. Rev. Lett.}
  \textbf{\bibinfo{volume}{86}}, \bibinfo{pages}{4791}.


\bibitem[{\citenamefont{Dukelsky \emph{et~al} }(2002)\citenamefont{Dukelsky, Esebbag and Pittel}}]{DEP}
\bibinfo{author}{\bibnamefont{Dukelsky}, \bibfnamefont{J.}},
  \bibinfo{author}{\bibfnamefont{C.}~\bibnamefont{Esebbag}}, and
  \bibinfo{author}{\bibfnamefont{S.}~\bibnamefont{Pittel}},
  \bibinfo{year}{2002}, \bibinfo{journal}{Phys. Rev. Lett.}
  \textbf{\bibinfo{volume}{88}}, \bibinfo{pages}{062501}.

\bibitem[{\citenamefont{Dukelsky \emph{et~al.}}(2003)\citenamefont{Dukelsky, Rom\'{a}n
  and Sierra}}]{Duke}
\bibinfo{author}{\bibnamefont{Dukelsky}, \bibfnamefont{J.}},
  \bibinfo{author}{\bibfnamefont{J.~M}~\bibnamefont{Rom\'an}}, and
  \bibinfo{author}{\bibfnamefont{G.}~\bibnamefont{Sierra}},
  \bibinfo{year}{2003}, \bibinfo{journal}{Phys. Rev. Lett.}
  \textbf{\bibinfo{volume}{90}}, \bibinfo{pages}{249803}.



\bibitem[{\citenamefont{Elliott}(1958)}]{El}
\bibinfo{author}{\bibnamefont{Elliott}, \bibfnamefont{J.~P.}},
  \bibinfo{year}{1958}, \bibinfo{journal}{Proc. Roc. Soc.}
  \textbf{\bibinfo{volume}{A242}}, \bibinfo{pages}{128, 562}.

\bibitem[{\citenamefont{Gaudin}(1976)}]{Gaudin}
\bibinfo{author}{\bibnamefont{Gaudin}, \bibfnamefont{M.}},
  \bibinfo{year}{1976}, \bibinfo{journal}{J. Phys. (Paris)}
  \textbf{\bibinfo{volume}{37}}, \bibinfo{pages}{1087}.

\bibitem[{\citenamefont{Gaudin}(1995)\citenamefont{Gaudin}}]{G-book}
\bibinfo{author}{\bibnamefont{Gaudin}, \bibfnamefont{M.}},
  \bibinfo{year}{1995}, \emph{\bibinfo{title}{\'Etats propres et valeurs propres de
l'Hamiltonien d'appariement}}
  (\bibinfo{publisher}{Les \'Editions de Physique}).

\bibitem[{\citenamefont{Gould \emph{et~al.}} (2002)\citenamefont{Gould, Zhang and Zhou}}]{Gould}
\bibinfo{author}{\bibnamefont{Gould}, \bibfnamefont{ M.~D.}},
  \bibinfo{author}{\bibfnamefont{Y.-Z.}~\bibnamefont{Zhang}}, and
  \bibinfo{author}{\bibfnamefont{S.-Y.}~\bibnamefont{Zhao}},
  \bibinfo{year}{2002}, \bibinfo{journal}{Nucl. Phys. B}
  \textbf{\bibinfo{volume}{630}}, \bibinfo{pages}{492}.

\bibitem[{\citenamefont{Greiner \emph{et~al.}}(2003)\citenamefont{Greiner, Regal and Jin}}]{GRJ}
\bibinfo{author}{\bibnamefont{Greiner}, \bibfnamefont{M.}},
  \bibinfo{author}{\bibfnamefont{C.~A.}~\bibnamefont{Regal}}, and
  \bibinfo{author}{\bibfnamefont{D.~S}~\bibnamefont{Jin}},
  \bibinfo{year}{2003}, \bibinfo{journal}{Naure (London)}
  \textbf{\bibinfo{volume}{426}}, \bibinfo{pages}{537}.

\bibitem[{\citenamefont{Guan \emph{et~al.}}(2002)\citenamefont{Guan, Foerster, Links and Zhou}}]{su4}
\bibinfo{author}{\bibfnamefont{Guan}, \bibnamefont{X.-W.}},
\bibinfo{author}{\bibfnamefont{A.}~\bibnamefont{Foerster}},
\bibinfo{author}{\bibfnamefont{ J.}~\bibnamefont{Links}}, and
  \bibinfo{author}{\bibfnamefont{H.-Q.}~\bibnamefont{Zhou}}
  \bibinfo{year}{2002}, \bibinfo{journal}{Nucl. Phys. B}
  \textbf{\bibinfo{volume}{642}}, \bibinfo{pages}{501}.

\bibitem[{\citenamefont{Ha}(1996)\citenamefont{Ha}}]{Ha}
\bibinfo{author}{\bibnamefont{Ha}, \bibfnamefont{Z.~N.~C.}},
  \bibinfo{year}{1996}, \emph{\bibinfo{booktitle}{Quantum Many-Body Systems in One Dimension}},
(\bibinfo{publisher}{World Scientific}).

\bibitem[{\citenamefont{Haldane}(1981)}]{Hal}
\bibinfo{author}{\bibnamefont{Haldane}, \bibfnamefont{F.~D.~M.}}, \bibinfo{year}{1981},
  \bibinfo{journal}{J. Phys. C} \textbf{\bibinfo{volume}{14}},
  \bibinfo{pages}{2585}.

\bibitem[{\citenamefont{Haldane}(1988)}]{Hal2}
\bibinfo{author}{\bibnamefont{Haldane}, \bibfnamefont{F.~D.~M.}}, \bibinfo{year}{1988},
  \bibinfo{journal}{Phys. Rev. Lett.} \textbf{\bibinfo{volume}{60}},
  \bibinfo{pages}{635}.


\bibitem[{\citenamefont{Hasegawa and Tasaki}(1987)}]{S2}
\bibinfo{author}{\bibnamefont{Hasegawa}, \bibfnamefont{M.}},
 and  \bibinfo{author}{\bibfnamefont{S.}~\bibnamefont{Tasaki}},
  \bibinfo{year}{1987}, \bibinfo{journal}{Phys. Rev. C}
  \textbf{\bibinfo{volume}{35}}, \bibinfo{pages}{1508}.

\bibitem[{\citenamefont{Hasegawa and Tasaki}(1993)}]{S3}
\bibinfo{author}{\bibnamefont{Hasegawa}, \bibfnamefont{M.}},
 and  \bibinfo{author}{\bibfnamefont{S.}~\bibnamefont{Tasaki}},
  \bibinfo{year}{1993}, \bibinfo{journal}{Phys. Rev. C}
  \textbf{\bibinfo{volume}{47}}, \bibinfo{pages}{188}.


\bibitem[{\citenamefont{Heiselberg}(2003)}]{Hei}
\bibinfo{author}{\bibnamefont{Heiselberg}, \bibfnamefont{H.}}, \bibinfo{year}{2003},
  \bibinfo{journal}{Phys. Rev. A} \textbf{\bibinfo{volume}{68}},
  \bibinfo{pages}{053616}.


\bibitem[{\citenamefont{Heritier}(2001)}]{He}
\bibinfo{author}{\bibnamefont{Heritier}, \bibfnamefont{M.}}, \bibinfo{year}{2001},
  \bibinfo{journal}{Nature (London)} \textbf{\bibinfo{volume}{414}},
  \bibinfo{pages}{31}.


\bibitem[{\citenamefont{Iachello and Arima}(1980)}]{Ia}
\bibinfo{author}{\bibnamefont{Iachello}, \bibfnamefont{F.}},
and \bibinfo{author}{\bibfnamefont{A.}~\bibnamefont{Arima}},
  \bibinfo{year}{1980}, \emph{\bibinfo{title}{The Interacting Boson Model}}
  (\bibinfo{publisher}{Cambridge University Press}).

\bibitem[{\citenamefont{Jochim} \emph{et~al.} (2003)\citenamefont{Jochim \emph{et~al.}}}]{JEA}
\bibinfo{author}{\bibnamefont{Jochim}, \bibfnamefont{ S.}},
  \bibinfo{author}{\bibfnamefont{\emph{et~al.}}},
  \bibinfo{journal}{Science}
  \textbf{\bibinfo{volume}{302}}, \bibinfo{pages}{5653}.


\bibitem[{\citenamefont{Links \emph{et~al.}}(2002)\citenamefont{Links, Zhou, Gould and
McKenzie}}]{o5}
\bibinfo{author}{\bibnamefont{Links}, \bibfnamefont{ J.}},
  \bibinfo{author}{\bibfnamefont{H.-Q.}~\bibnamefont{Zhou}},
  \bibinfo{author}{\bibfnamefont{M.~D.}~\bibnamefont{Gould}}, and \bibinfo{author}
  {\bibfnamefont{R.~H.}~\bibnamefont{McKenzie}}
  \bibinfo{year}{2002}, \bibinfo{journal}{J. Phys. A}
  \textbf{\bibinfo{volume}{35}}, \bibinfo{pages}{6459}.


\bibitem[{\citenamefont{ Links \emph{et~al.}}(2003)\citenamefont{ Links, Zhou, McKenzie
  and Gould}}]{Lin}
\bibinfo{author}{\bibnamefont{Links}, \bibfnamefont{ J.}},
  \bibinfo{author}{\bibfnamefont{H.-Q.}~\bibnamefont{Zhou}},
  \bibinfo{author}{\bibfnamefont{R.~H.}~\bibnamefont{McKenzie}},  and
  \bibinfo{author}{\bibfnamefont{M.~D.}~\bibnamefont{Gould}},
  \bibinfo{year}{2003}, \bibinfo{journal}{J. Phys. A}
  \textbf{\bibinfo{volume}{36}}, \bibinfo{pages}{R63}.



\bibitem[{\citenamefont{Luttinger}(1963)}]{Lut}
\bibinfo{author}{\bibnamefont{Luttinger}, \bibfnamefont{J.~M.}}, \bibinfo{year}{1963},
  \bibinfo{journal}{J. Math. Phys.} \textbf{\bibinfo{volume}{15}},
  \bibinfo{pages}{609}.



\bibitem[{\citenamefont{Mastellone \emph{et~al.}}(1998)\citenamefont{Mastellone, Falci
  and Fazio}}]{Lanczos}
\bibinfo{author}{\bibnamefont{Mastellone}, \bibfnamefont{A.}},
  \bibinfo{author}{\bibfnamefont{G.}~\bibnamefont{Falci}}, and
  \bibinfo{author}{\bibfnamefont{R.}~\bibnamefont{Fazio}},
  \bibinfo{year}{1998}, \bibinfo{journal}{Phys. Rev. Lett.}
  \textbf{\bibinfo{volume}{80}}, \bibinfo{pages}{4542}.


\bibitem[{\citenamefont{Matveev and Larkin}(1997)}]{Matveev}
\bibinfo{author}{\bibnamefont{Matveev}, \bibfnamefont{K.~A.}}, and
  \bibinfo{author}{\bibfnamefont{A.~I.}~\bibnamefont{Larkin}},
  \bibinfo{year}{1997}, \bibinfo{journal}{Phys. Rev. Lett.}
  \textbf{\bibinfo{volume}{78}}, \bibinfo{pages}{3749}.


\bibitem[{\citenamefont{Nozieres and Saint James}(1982)}]{NSJ}
\bibinfo{author}{\bibnamefont{Nozieres}, \bibfnamefont{P.}}, and
  \bibinfo{author}{\bibfnamefont{D.}~\bibnamefont{Saint James}},
  \bibinfo{year}{1982}, \bibinfo{journal}{J. Phys. (Paris)}
  \textbf{\bibinfo{volume}{43}}, \bibinfo{pages}{1133}.

\bibitem[{\citenamefont{Pan and Draayer}(2002)}]{Pan}
\bibinfo{author}{\bibnamefont{Pan}, \bibfnamefont{F.}}, and
  \bibinfo{author}{\bibfnamefont{J.~P.}~\bibnamefont{Draayer}},
  \bibinfo{year}{2002}, \bibinfo{journal}{Phys. Rev. C}
  \textbf{\bibinfo{volume}{66}}, \bibinfo{pages}{044314}.

\bibitem[{\citenamefont{Pittel and Dukelsky}(2003)}]{DP2}
\bibinfo{author}{\bibnamefont{Pittel}, \bibfnamefont{S.}}, and
  \bibinfo{author}{\bibfnamefont{J.}~\bibnamefont{Dukelsky}},
  \bibinfo{year}{2003}, \eprint{nucl-th/0309020}.



\bibitem[{\citenamefont{Ralph, Black
  and Tinkham}(1997)\citenamefont{Ralph \emph{et~al.}}}]{RBT}
\bibinfo{author}{\bibnamefont{Ralph}, \bibfnamefont{D.~C.}},
  \bibinfo{author}{\bibfnamefont{C.~T.}~\bibnamefont{Black}}, and
  \bibinfo{author}{\bibfnamefont{M.}~\bibnamefont{Tinkham}},
  \bibinfo{year}{1997}, \bibinfo{journal}{Phys. Rev. Lett.}
  \textbf{\bibinfo{volume}{78}}, \bibinfo{pages}{4087}.


\bibitem[{\citenamefont{Regal \emph{et~al.}}(2004)\citenamefont{Regal, Greiner and Jin}}]{RGJ}
\bibinfo{author}{\bibnamefont{Regal}, \bibfnamefont{C.~A.}},
  \bibinfo{author}{\bibfnamefont{M.}~\bibnamefont{Greiner}}, and
  \bibinfo{author}{\bibfnamefont{D.~S.}~\bibnamefont{Jin}},
  \bibinfo{year}{2004}, \bibinfo{journal}{Phys. Rev. Lett.}
  \textbf{\bibinfo{volume}{92}}, \bibinfo{pages}{040403}.

\bibitem[{\citenamefont{Rela\~{n}o \emph{et~al.}}(2004)\citenamefont{Rela\~{n}o,  Gomez, Retamosa and Dukelsky}}]{Rela}
\bibinfo{author}{\bibnamefont{Rela\~{n}o}, \bibfnamefont{A.}},
  \bibinfo{author}{\bibfnamefont{J.~M.~G.}~\bibnamefont{Gomez}},
  \bibinfo{author}{\bibfnamefont{J.}~\bibnamefont{Retamosa}}, and
  \bibinfo{author}{\bibfnamefont{J.}~\bibnamefont{Dukelsky}},
  \bibinfo{year}{2004}, \eprint{nlin.CD/0402011}.


\bibitem[{\citenamefont{Richardson}(1963a)}]{R1}
\bibinfo{author}{\bibnamefont{Richardson}, \bibfnamefont{R.~W.}}, \bibinfo{year}{1963a},
  \bibinfo{journal}{Phys. Lett.} \textbf{\bibinfo{volume}{3}},
  \bibinfo{pages}{277}.

\bibitem[{\citenamefont{Richardson}(1963b)}]{R2}
\bibinfo{author}{\bibnamefont{Richardson}, \bibfnamefont{R.~W.}}, \bibinfo{year}{1963b},
  \bibinfo{journal}{Phys. Lett.} \textbf{\bibinfo{volume}{5}},
  \bibinfo{pages}{82}.


\bibitem[{\citenamefont{Richardson and Sherman}(1964) }]{R3}
\bibinfo{author}{\bibnamefont{Richardson}, \bibfnamefont{R.~W.}}, and
\bibinfo{author}{\bibfnamefont{N.}~\bibnamefont{Sherman}},
 \bibinfo{year}{1964},
  \bibinfo{journal}{Nucl. Phys. B} \textbf{\bibinfo{volume}{52}},
  \bibinfo{pages}{221}.



\bibitem[{\citenamefont{Richardson}(1965)}]{R4}
\bibinfo{author}{\bibnamefont{Richardson}, \bibfnamefont{R.~W.}}, \bibinfo{year}{1965},
  \bibinfo{journal}{J. Math. Phys.} \textbf{\bibinfo{volume}{6}},
  \bibinfo{pages}{1034}.

\bibitem[{\citenamefont{Richardson}(1966a)}]{R5}
\bibinfo{author}{\bibnamefont{Richardson}, \bibfnamefont{R.~W.}}, \bibinfo{year}{1966a},
  \bibinfo{journal}{Phys. Rev.} \textbf{\bibinfo{volume}{141}},
  \bibinfo{pages}{949}.

\bibitem[{\citenamefont{Richardson}(1966b)}]{R8}
\bibinfo{author}{\bibnamefont{Richardson}, \bibfnamefont{R.~W.}}, \bibinfo{year}{1966b},
  \bibinfo{journal}{Phys. Rev.} \textbf{\bibinfo{volume}{144}},
  \bibinfo{pages}{874}.

\bibitem[{\citenamefont{Richardson}(1967)}]{R9}
\bibinfo{author}{\bibnamefont{Richardson}, \bibfnamefont{R.~W.}}, \bibinfo{year}{1967},
  \bibinfo{journal}{Phys. Rev.} \textbf{\bibinfo{volume}{159}},
  \bibinfo{pages}{792}.

\bibitem[{\citenamefont{Richardson}(1968)}]{R6}
\bibinfo{author}{\bibnamefont{Richardson}, \bibfnamefont{R.~W.}}, \bibinfo{year}{1968},
  \bibinfo{journal}{J. Math. Phys.} \textbf{\bibinfo{volume}{9}},
  \bibinfo{pages}{1327}.

\bibitem[{\citenamefont{Richardson}(1977)}]{R7}
\bibinfo{author}{\bibnamefont{Richardson}, \bibfnamefont{R.~W.}}, \bibinfo{year}{1977},
  \bibinfo{journal}{J. Math. Phys.} \textbf{\bibinfo{volume}{18}},
  \bibinfo{pages}{1802}.



\bibitem[{\citenamefont{Roman \emph{et~al.}}(2002)\citenamefont{Roman,  Sierra
  and Dukelsky}}]{RSD}
\bibinfo{author}{\bibnamefont{Roman}, \bibfnamefont{ J.~M}},
  \bibinfo{author}{\bibfnamefont{G.}~\bibnamefont{Sierra}},  and
  \bibinfo{author}{\bibfnamefont{J.}~\bibnamefont{Dukelsky}},
  \bibinfo{year}{2002}, \bibinfo{journal}{Nucl. Phys. B}
  \textbf{\bibinfo{volume}{634}}, \bibinfo{pages}{483}.


\bibitem[{\citenamefont{Roman \emph{et~al.}}(2003)\citenamefont{Roman,  Sierra
  and Dukelsky}}]{RSD2}
\bibinfo{author}{\bibnamefont{Roman}, \bibfnamefont{ J.~M}},
  \bibinfo{author}{\bibfnamefont{G.}~\bibnamefont{Sierra}},  and
  \bibinfo{author}{\bibfnamefont{J.}~\bibnamefont{Dukelsky}},
  \bibinfo{year}{2003}, \bibinfo{journal}{Phys. Rev. B}
  \textbf{\bibinfo{volume}{67}}, \bibinfo{pages}{ 064510}.

\bibitem[{\citenamefont{Rombouts \emph{et~al.}}(2003)\citenamefont{Rombouts,  Van Neck
  and Dukelsky}}]{Romb}
\bibinfo{author}{\bibnamefont{Rombouts}, \bibfnamefont{ S.}},
  \bibinfo{author}{\bibfnamefont{D.}~\bibnamefont{Van Neck}},  and
  \bibinfo{author}{\bibfnamefont{J.}~\bibnamefont{Dukelsky}},
 \bibinfo{year}{2003}, \eprint{nuc-th/0312070}.


\bibitem[{\citenamefont{Schechter \emph{et~al.}}(2001)\citenamefont{Schechter, Imry, Levinson
  and Gould}}]{Sch}
\bibinfo{author}{\bibnamefont{Schechter}, \bibfnamefont{ M.}},
  \bibinfo{author}{\bibfnamefont{Y.}~\bibnamefont{Imry}},
  \bibinfo{author}{\bibfnamefont{Y.}~\bibnamefont{Levinson}},  and
  \bibinfo{author}{\bibfnamefont{J.}~\bibnamefont{von Delft}},
  \bibinfo{year}{2001}, \bibinfo{journal}{Phys. Rev. B}
  \textbf{\bibinfo{volume}{63}}, \bibinfo{pages}{214518}.



\bibitem[{\citenamefont{Shastry}(1988)}]{Shas}
\bibinfo{author}{\bibnamefont{Shastry}, \bibfnamefont{B.~S.}}, \bibinfo{year}{1988},
  \bibinfo{journal}{Phys. Rev. Lett.} \textbf{\bibinfo{volume}{60}},
  \bibinfo{pages}{639}.

\bibitem[{\citenamefont{Shastry and Dhar}(2001)}]{Shas2}
\bibinfo{author}{\bibnamefont{Shastry}, \bibfnamefont{B.~S.}},
  and   \bibinfo{author}{\bibfnamefont{A.}~\bibnamefont{Dhar}},
  \bibinfo{year}{2001}, \bibinfo{journal}{J. Phys. A}
  \textbf{\bibinfo{volume}{34}}, \bibinfo{pages}{6197}.



\bibitem[{\citenamefont{Sierra}(2000)}]{Si}
\bibinfo{author}{\bibnamefont{Sierra}, \bibfnamefont{G.}}, \bibinfo{year}{2000},
  \bibinfo{journal}{Nucl. Phys. B} \textbf{\bibinfo{volume}{572}},
  \bibinfo{pages}{517}.



\bibitem[{\citenamefont{Sierra \emph{et~al.}}(2000)
\citenamefont{Sierra, Dukelsky, Dussel, von Delft and
Braun}}]{random}
\bibinfo{author}{\bibnamefont{Sierra}, \bibfnamefont{G.}},
  \bibinfo{author}{\bibfnamefont{J.}~\bibnamefont{Dukelsky}},
  \bibinfo{author}{\bibfnamefont{G.~G.}~\bibnamefont{Dussel}},
  \bibinfo{author}{\bibfnamefont{J.}~\bibnamefont{von Delft}},and
  \bibinfo{author}{\bibfnamefont{F.}~\bibnamefont{Braun}},
  \bibinfo{year}{2000}, \bibinfo{journal}{Phys. Rev. B}
  \textbf{\bibinfo{volume}{ 61}}, \bibinfo{pages}{11890}.


\bibitem[{\citenamefont{Sierra}}(2001)]{Scomo}
\bibinfo{author}{\bibnamefont{Sierra}, \bibfnamefont{G.}},
  \bibinfo{year}{2001}, \eprint{hep-th/0111114}.


\bibitem[{\citenamefont{Sigrist and Ueda}(1991)}]{SU}
\bibinfo{author}{\bibnamefont{Sigrist}, \bibfnamefont{M.}},
  and   \bibinfo{author}{\bibfnamefont{K.}~\bibnamefont{Ueda}},
  \bibinfo{year}{1991}, \bibinfo{journal}{Rev. Mod. Phys.}
  \textbf{\bibinfo{volume}{63}}, \bibinfo{pages}{239}.



\bibitem[{\citenamefont{Smith and Ambegaokar}(1996)}]{SA}
\bibinfo{author}{\bibnamefont{Smith}, \bibfnamefont{R.~A.}},
  and   \bibinfo{author}{\bibfnamefont{V.}~\bibnamefont{Ambegaokar}},
  \bibinfo{year}{1996}, \bibinfo{journal}{Phys. Rev. Lett.}
  \textbf{\bibinfo{volume}{77}}, \bibinfo{pages}{4962}.



\bibitem[{\citenamefont{Sutherland}(1971)}]{Sut}
\bibinfo{author}{\bibnamefont{Sutherland}, \bibfnamefont{B.}}, \bibinfo{year}{1971},
  \bibinfo{journal}{J. Math. Phys.} \textbf{\bibinfo{volume}{12}},
  \bibinfo{pages}{246}.


\bibitem[{\citenamefont{Tomonaga}(1950)}]{Tomo}
\bibinfo{author}{\bibnamefont{Tomonaga}, \bibfnamefont{S.}}, \bibinfo{year}{1950},
  \bibinfo{journal}{Prog. Theor. Phys.} \textbf{\bibinfo{volume}{5}},
  \bibinfo{pages}{544}.

\bibitem[{\citenamefont{Volya \emph{et~al.}}(2002)\citenamefont{Volya, Zelevinsky and Brown}}]{VZB}
\bibinfo{author}{\bibnamefont{Volya}, \bibfnamefont{A.}},
  \bibinfo{author}{\bibfnamefont{V.}~\bibnamefont{Zelevinsky}},  and
  \bibinfo{author}{\bibfnamefont{B. A.}~\bibnamefont{Brown}},
  \bibinfo{year}{2002}, \bibinfo{journal}{Phys. Rev. C}
  \textbf{\bibinfo{volume}{65}}, \bibinfo{pages}{054312}.


\bibitem[{\citenamefont{von Delft \emph{et~al.}}(1996)\citenamefont{von Delft, Golubev, Tichy
and Zaikin,}}]{grains}
\bibinfo{author}{\bibnamefont{von Delft}, \bibfnamefont{ J.}},
  \bibinfo{author}{\bibfnamefont{D.~S.}~\bibnamefont{Golubev}},
  \bibinfo{author}{\bibfnamefont{W.}~\bibnamefont{Tichy}},  and
  \bibinfo{author}{\bibfnamefont{A.~D.}~\bibnamefont{Zaikin}},
  \bibinfo{year}{1996}, \bibinfo{journal}{Phys. Rev. Lett.}
  \textbf{\bibinfo{volume}{77}}, \bibinfo{pages}{3189}.

\bibitem[{\citenamefont{von Delft and Ralph}(2001)}]{Rep}
\bibinfo{author}{\bibnamefont{von Delft}, \bibfnamefont{J.}},
  and   \bibinfo{author}{\bibfnamefont{D.~C.}~\bibnamefont{Ralph}},
  \bibinfo{year}{2001}, \bibinfo{journal}{Phys. Rep.}
  \textbf{\bibinfo{volume}{345}}, \bibinfo{pages}{61}.



\bibitem[{\citenamefont{von Delft and Poghossian}(2002)}]{Pog}
\bibinfo{author}{\bibnamefont{von Delft}, \bibfnamefont{J.}}, and
  \bibinfo{author}{\bibfnamefont{R.}~\bibnamefont{Poghossian}},
  \bibinfo{year}{2002}, \bibinfo{journal}{Phys. Rev. B}
  \textbf{\bibinfo{volume}{66}}, \bibinfo{pages}{134502}.


\bibitem[{\citenamefont{Wu \emph{et~al.}}(2003) \citenamefont{Wu, Guidry, Sun
  and Wu}}]{Wu}
\bibinfo{author}{\bibnamefont{Wu}, \bibfnamefont{ L.-A.}},
  \bibinfo{author}{\bibfnamefont{M.~W.}~\bibnamefont{Guidry}},
  \bibinfo{author}{\bibfnamefont{Y.}~\bibnamefont{Sun}},  and
  \bibinfo{author}{\bibfnamefont{C.-L.}~\bibnamefont{Wu}},
  \bibinfo{year}{2002}, \bibinfo{journal}{Phys. Rev. B}
  \textbf{\bibinfo{volume}{67}}, \bibinfo{pages}{014515}.

\bibitem[{\citenamefont{Yuzbashyan \emph{et~al.}}(2003)
\citenamefont{Yuzbashyan, Baytin and Altshuler}}]{Yuz}
\bibinfo{author}{\bibnamefont{Yuzbashyan}, \bibfnamefont{E.~A.}},
\bibinfo{author}{\bibfnamefont{A.~A.}~\bibnamefont{Baytin}},  and
  \bibinfo{author}{\bibfnamefont{B.~L.}~\bibnamefont{Altshuler}},
  \bibinfo{year}{2003}, \eprint{cond-mat/0305635}.


\bibitem[{\citenamefont{Zhang} (1997)}]{Zhang}
\bibinfo{author}{\bibnamefont{Zhang}, \bibfnamefont{ S.-C.}},
  \bibinfo{year}{1997}, \bibinfo{journal}{Science}
  \textbf{\bibinfo{volume}{275}}, \bibinfo{pages}{1089}.

\bibitem[{\citenamefont{Zhou \emph{et~al.}}(2002)\citenamefont{Zhou, Links, McKenzie
  and Gould}}]{ZLMG}
\bibinfo{author}{\bibnamefont{Zhou}, \bibfnamefont{ H.-Q.}},
  \bibinfo{author}{\bibfnamefont{J.}~\bibnamefont{Links}},
  \bibinfo{author}{\bibfnamefont{R.~H.}~\bibnamefont{McKenzie}},  and
  \bibinfo{author}{\bibfnamefont{M.~D.}~\bibnamefont{Gould}},
  \bibinfo{year}{2002}, \bibinfo{journal}{Phys. Rev. B}
  \textbf{\bibinfo{volume}{65}}, \bibinfo{pages}{060502}.

\end{thebibliography}

\end{document}